%% file: main.tex
\documentclass[11pt]{article}

\usepackage[numbers]{natbib}
\usepackage{amsmath}
\usepackage{amssymb}
\usepackage{amsthm}
\usepackage{xcolor}
\usepackage{array}
\usepackage{booktabs}
\usepackage{tabularx}
\usepackage{hyperref}
\usepackage{algorithm}
\usepackage{algpseudocode}
\usepackage[margin=1in]{geometry}

\hypersetup{colorlinks=true,citecolor=blue,linkcolor=blue,hypertexnames=false}
\allowdisplaybreaks

\input{_0_vars}

\input{_1_maths}

\theoremstyle{plain}
\newtheorem{theorem}{Theorem}[section]
\newtheorem{lemma}[theorem]{Lemma}

\theoremstyle{definition}
\newtheorem{definition}[theorem]{Definition}

\begin{document}

\date{September 29, 2026}
\title{\paperTitle}
\author{\paperAuthor}
\maketitle
\begin{abstract}
\input{00_abstract}

\end{abstract}

\newpage
\input{_2_body}

\bibliographystyle{alpha}
\bibliography{ref}

\appendix

\input{_3_app}

\end{document}

%% file: _0_vars.tex
\newcommand{\paperTitle}{Fixed-Depth Bootstrapping for Directed Shortest Paths}
\newcommand{\paperAuthor}{Jerry Yao-Chieh Hu\thanks{Vanderbilt University. \texttt{jhu@u.northwestern.edu}.} \and Zhao Song\thanks{\texttt{magic.linuxkde@gmail.com}. The author would like to thank Josh Alman, Brayden Letwin, George Li, Victor Reis, Omri Weinstein, Lichen Zhang, and Ruizhe Zhang for very helpful discussions.}
}

%% file: _1_maths.tex
\newcommand{\R}{\mathbb R}

%% file: 00_abstract.tex
The single-source shortest path problem asks for the exact distance from a given source to every vertex in a graph. 
For directed graphs with $n$ vertices, $m$ edges, and nonnegative real edge weights, we give a deterministic algorithm that runs in $O(n+m\sqrt{\log n}(\log\log n)^{1/4})$ time in the comparison-addition model.
On graphs with $m=\Theta(n)$, our running time bound improves upon that of Duan, Mao, Shu, and Yin~\cite{dmsy26} by a factor of $(\log\log n)^{1/4}$.

%% file: _2_body.tex
\input{01_intro}

\input{20_preli}
\input{30_upper_bound}

%% file: 01_intro.tex
\section{Introduction}

Given a graph with $n$ vertices, $m$ edges, and a source $s$, the single-source shortest-path (SSSP) problem asks for the minimum total weight of a path from $s$ to each vertex. 
Following Duan, Mao, Shu, and Yin~\cite{dmsy26}, we study directed graphs with nonnegative real edge weights in the comparison-addition model~\cite{pr05}, in which the only operations on weights are addition
and comparison, each taking constant time.

We first state prior bounds for the source-reachable graph:
\begin{itemize}
    \item The Bellman-Ford algorithm~\cite{ford56,bellman58} allows negative weights and runs in $O(mn)$ time when no negative cycle is reachable from $s$. 
    Bernstein, Nanongkai, and Wulff-Nilsen~\cite{bnw22} obtained a near-linear bound for polynomially bounded integer weights of either sign, later improved by Bringmann, Cassis, and Fischer~\cite{bcf23}.
    For real weights of either sign, Fineman~\cite{fin24} improved on the Bellman-Ford bound. 
    Subsequent work used hop reduction~\cite{hjq26} and sparsification~\cite{qt26}, and Hair et al.~\cite{hllz26} obtained an $m^{1+o(1)}$-time algorithm.

    \item For nonnegative weights, Dijkstra's algorithm~\cite{dijkstra59} runs in $O(m+n\log n)$ time with the Fibonacci heaps of Fredman and Tarjan~\cite{ft87}. Dijkstra also orders the reachable vertices by distance. Haeupler et al.~\cite{hhrrt24} showed that, with a suitable heap, Dijkstra's algorithm is universally optimal for computing this order. SSSP asks only for the distances.

    \item For nonnegative integer weights, Ahuja et al.~\cite{amot90} developed radix heaps for Dijkstra's algorithm. Fredman and Willard~\cite{fw94} gave an $O(m+n\log n/\log\log n)$ implementation on the word RAM. Thorup~\cite{thorup99} gave a deterministic linear-time algorithm for undirected graphs with positive integer weights. His later integer priority queues improved the directed bound~\cite{thorup04}.

    \item 
    For undirected graphs with positive real weights, Pettie and Ramachandran~\cite{pr05} gave a bound that depends on the ratio of the largest to the smallest weight. For arbitrary nonnegative real weights, Duan, Mao, Shu, and Yin~\cite{dmsy23} gave a randomized $O(m\sqrt{\log n\,\log\log n})$-time algorithm. Yan~\cite{yan26} obtained the same bound deterministically. Kadria and Roditty~\cite{kr26} improved the randomized bound to $O(m\sqrt{\log n}\,(\log\log n\,\log\log\log n)^{1/4})$.

    \item For directed graphs with nonnegative real weights, Duan, Mao, Mao, Shu, and Yin~\cite{dmmsy25} gave a deterministic $O(m\log^{2/3} n)$-time algorithm, breaking the sorting barrier when $m=\Theta(n)$. Duan, Mao, Shu, and Yin~\cite{dmsy26} improved the bound to $O(m\sqrt{\log n}+\sqrt{mn\log n\,\log\log n})$. For $m=\Theta(n)$, this is $O(m\sqrt{\log n\,\log\log n})$.

\end{itemize}

\begin{table}[htbp]
\centering
\footnotesize

\caption{Selected exact SSSP bounds. Big-$O$ notation is omitted. D/U denotes directed/undirected in the Graph column, R/D denotes randomized/deterministic in the R/D column, CA denotes comparison-addition, and RAM denotes word-sized integer weights. The bounds concern the source-reachable graph. General inputs require $O(n+m)$ preprocessing. The randomized bounds hold with high probability. Here $C$ is the maximum integer weight, $r$ is the ratio of maximum to minimum positive weight, and $\alpha$ is the inverse-Ackermann function. Logarithmic factors are truncated below by $1$. Bellman-Ford \cite{ford56,bellman58} assumes no reachable negative cycle.}
\label{tab:sssp_history}
\begin{tabular}{|l|l|l|l|l|l|l|l|}
\toprule
Year & Authors & References & Graph & Weights & Model & R/D & Running time \\
\midrule
1956 & Ford, Bellman & \cite{ford56,bellman58}
& D & $\R$ & CA & D
& $mn$ \\
\hline
1959 & Dijkstra & \cite{dijkstra59}
& D/U & $\R_{\geq 0}$ & CA & D
& $n^2+m$ \\
\hline
1987 & Fredman-Tarjan & \cite{ft87}
& D/U & $\R_{\geq 0}$ & CA & D
& $m+n\log n$ \\
\hline
1990 & Ahuja-Mehlhorn-Orlin-Tarjan & \cite{amot90}
& D & $\mathbb{N}_{\geq 0}$ & RAM & D
& $m+n\sqrt{\log C}$ \\
\hline
1994 & Fredman-Willard & \cite{fw94}
& D & $\mathbb{N}_{\geq 0}$ & RAM & D
& $m+\frac{n\log n}{\log_2 n}$ \\
\hline
1999 & Thorup & \cite{thorup99}
& U & $\mathbb{N}_{>0}$ & RAM & D
& $m+n$ \\
\hline
2004 & Thorup & \cite{thorup04}
& D & $\mathbb{N}_{\geq 0}$ & RAM & D
& $m+n\log_2\min\{n,C\}$ \\
\hline
2005 & Pettie-Ramachandran & \cite{pr05}
& U & $\R_{>0}$ & CA & D
& $m\alpha(m,n)+\min\{n\log n,n\log_2 r\}$ \\
\hline
2023 & Duan-Mao-Shu-Yin & \cite{dmsy23}
& U & $\R_{\geq 0}$ & CA & R
& $m\sqrt{\log n\log_2 n}$ \\
\hline
2025 & Duan-Mao-Mao-Shu-Yin & \cite{dmmsy25}
& D & $\R_{\geq 0}$ & CA & D
& $m\log^{2/3}n$ \\
\hline
2026 & Yan & \cite{yan26}
& U & $\R_{\geq 0}$ & CA & D
& $m\sqrt{\log n\log_2 n}$ \\
\hline
2026 & Duan-Mao-Shu-Yin & \cite{dmsy26}
& D & $\R_{\geq 0}$ & CA & D
& $m\sqrt{\log n}+\sqrt{mn\log n\log_2 n}$ \\
\hline
2026 & Kadria-Roditty & \cite{kr26}
& U & $\R_{\geq 0}$ & CA & R
& $m\sqrt{\log n}(\log_2 n)^{1/4}$
  $(\log_3 n)^{1/4}$ \\
\hline
2026 & This work & Theorem~\ref{thm:informal_main} 
& D & $\R_{\geq 0}$ & CA & D
& $m\sqrt{\log n}(\log_2 n)^{1/4}$
  \\ \hline

\end{tabular}

\end{table}

\subsection{Our Results}
\label{sec:our_results}

We state our main result as follows.
Throughout the paper, we follow the notation of Duan, Mao, Shu, and Yin~\cite{dmsy26}.
We write $\log_j n$ for the $j$-th iterated natural logarithm.

\begin{theorem}[Informal version of Theorem~\ref{thm:short_walk_solver}]
\label{thm:informal_main}
Given a directed graph with $n$ vertices, $m$ edges, nonnegative real edge weights, and a source vertex, one can deterministically compute all exact distances from the source in the comparison-addition model in time
$
O(n+m\sqrt{\log n}(\log_2 n)^{1/4}).
$
\end{theorem}
For the situation $m = o( n (\log_2 n)^{1/2})$, our result improves \cite{dmsy26}.
Intuitively, Duan, Mao, Mao, Shu, and Yin~\cite{dmsy26} use bounded searches to avoid ordering all vertices by distance. We use a second bounded solver for the local Dijkstra searches that group the frontier vertices. At the smallest scale, searches from the same root reuse a precomputed order of short walks.

%% file: 20_preli.tex
\section{Preliminaries}
\label{sec:preliminaries}

We first define notation for asymptotic bounds, graphs, and distances. We then give our conventions for logarithms and local instance sizes and specify which operations the comparison-addition model allows.

\paragraph{Numbers and asymptotic notation.}
We write $\R$ for the real numbers, $\R_{>0}$ and $\R_{\geq 0}$ for the positive and nonnegative reals, and $\mathbb{N}_{>0}$ and $\mathbb{N}_{\geq 0}$ for the positive and nonnegative integers. For a positive integer $q$, $[q]:=\{1,\ldots,q\}$. We use $O$, $\Omega$, and $\Theta$ for asymptotic upper, lower, and matching bounds, and $f=o(g)$ when $\frac{f}{g}\to0$. A subscript on an asymptotic bound permits its constant to depend on that parameter. Unless stated otherwise, implied constants are absolute. We use $\infty$ for an unavailable distance or label and set $\min\varnothing:=\infty$.

\paragraph{Graphs, walks, and distances.}
Let $G:=(V,E)$ be a directed graph with source $s\in V$ and edge-weight function $w:E\to\R_{\geq 0}$. Let $n:=|V|$ and $m:=|E|$. A directed walk $P$ has hop count equal to its number of edges, and weight $w(P)$ equal to the sum of their weights, counting repetitions. A simple path has no repeated vertex. The empty path has weight and hop count zero. We write $\operatorname{dist}_G(u,v)$ for the minimum weight of a directed $u$-to-$v$ path, or $\infty$ if none exists. The required output is $\operatorname{dist}_G(s,v)$ for every $v\in V$. A vertex is source-reachable if this value is finite. General-input bounds include the $O(n)$ initialization and output cost. Restricting to source-reachable vertices takes $O(n+m)$ time. A reachable graph has $m\ge n-1$.

\paragraph{Subgraphs and degrees.}
For $X\subseteq V$, $G[X]$ is the induced directed subgraph, and $E(X)$ is the set of edges with both endpoints in $X$, with the ambient graph specified by context. We write $\deg_G^+(v)$ and $\deg_G^-(v)$ for the out-degree and in-degree of $v$, and omit $G$ when it is clear. A bound $\Delta$ on the degree bounds both in-degree and out-degree unless specified otherwise. A simple directed graph has no self-loop and at most one edge for each ordered pair of vertices. Opposite directed edges are allowed. Parallel edges can be coalesced by keeping a minimum-weight edge for each ordered pair, and nonnegative self-loops can be removed.

\paragraph{Trees.}
In a rooted tree, ancestors and descendants include the vertex itself unless called proper.

\paragraph{Logarithms and growth functions.}
The symbol $\log$ denotes the natural logarithm and $\lg$ denotes the base-two logarithm. Let $\log_0x:=x$ and $\log_{j+1}x:=\log(\log_jx)$ whenever defined. Thus $\log_2x=\log\log x$, whereas $\log^2x=(\log x)^2$. In asymptotic bounds, each logarithmic factor is truncated below by a fixed positive constant. Recursive definitions use their stated base cases. For positive integers $q$, let $\beta(q):=\max\{4,\lceil\lg q\rceil\}$.

\paragraph{Local instance sizes.}
For a local graph instance, $N$ and $M_G$ denote its numbers of vertices and edges, respectively, and $Q\ge N$ is an advertised upper bound on its number of vertices.

\paragraph{Computational model.}
Edge weights are accessed through comparisons and additions. Vertex identifiers, pointers, ranks, and counters use ordinary words of $O(\log(n+m))$ bits in the ambient instance. After coalescing, $m\le n^2$, so $O(\log n)$ bits suffice. The algorithm may perform ordinary word operations on this integer metadata, but does not inspect bits of real weights. Auxiliary and restricted instances retain the ambient identifier convention. Statements about local charged work exclude reusable ambient initialization only when this is stated explicitly. Standalone running-time bounds include initialization and output.

%% file: 30_upper_bound.tex
\section{Proof of Theorem~\ref{thm:informal_main}}
\label{sec:upper_bound}

Section~\ref{sec:path_labels} defines path labels and constant-time relaxation tests. Section~\ref{sec:capped_feasibility} proves that shared-label updates and cap tightening preserve the local problem. Section~\ref{sec:frontier_certificates} establishes frontier certificates and cumulative discovery-tree accounting. Section~\ref{sec:capped_engine} specifies the recursive capped engine and its saved-threshold invariant. Section~\ref{sec:inherited_histories} handles inherited histories, nested environments, and small completions of execution prefixes. Section~\ref{sec:engine_correctness} proves the capped return contract. Section~\ref{sec:queue_provenance} proves the queue and provenance bounds, and Section~\ref{sec:parameterized_accounting} combines them into a parameterized cost bound. Section~\ref{sec:first_bootstrap} converts complete bounds into interruption bounds. Finally, Section~\ref{sec:walk_rank_refinement} precomputes orders for short walks and proves the formal version of Theorem~\ref{thm:informal_main}.

\subsection{Path labels and constant-time relaxation tests}
\label{sec:path_labels}

Initially consider a simple directed graph of constant maximum in-degree and out-degree. A standard degree reduction introduces zero-weight edges and $O(n+m)$ vertices and edges. After restricting to the source-reachable subgraph, the expanded size is $O(m)$. Parallel edges can first be coalesced by retaining a minimum-weight edge for each ordered pair of endpoints, using endpoint stamps while scanning each adjacency list. Nonnegative self-loops can be removed.

Zero-weight edges allow paths with different numbers of edges to have the same weight. To choose a canonical path, we break ties first by hop count and then by reading the vertex identifiers backwards.

\begin{definition}[Path labels and accepted relaxations]
\label{def:path_labels}
For a path $P:=(v_0,\ldots,v_h)$, define its mathematical label and scheduling key by
\[
L(P):=(w(P),h,v_h,v_{h-1},\ldots,v_0),\qquad \rho(L(P)):=(w(P),h,v_h).
\]
Labels are ordered lexicographically. The engine stores only the fields specified below. Full reverse vertex sequences define its mathematical labels. Extending a path by an edge strictly increases its scheduling key: either the weight increases or the hop count increases. A cutoff is a scheduling key or infinity, and an arrival is below a cutoff $B$ when its key is strictly less than $B$.

Comparisons between different endpoints and comparisons with cutoffs require only the three-field scheduling keys.

Maintain a shared label $\lambda(v)$ that is the label of a genuine path from the original source, or infinity. Each accepted relaxation does not increase this full mathematical label. The stored representation consists of its weight, hop count, endpoint, and predecessor. An accepted relaxation includes equality. An equality relaxation can still cause first admission to a local search.
\end{definition}

The full label of a long path contains many vertex identifiers. A relaxation only compares a fresh extension with the path already stored at the same endpoint, so we can make this comparison from four fields.

\begin{lemma}[Four-field relaxation test]
\label{lem:four_field_test}
Use the labels and accepted relaxations of Definition~\ref{def:path_labels}. Suppose they evolve monotonically in a simple graph. Whether a fresh extension of the current $\lambda(u)$ along $(u,v)$ is at most the current $\lambda(v)$ can be decided from the four stored fields.
\end{lemma}
\begin{proof}
We use the lexicographic order of Definition~\ref{def:path_labels} to compare the fresh extension with the stored label at their common endpoint $v$. Their endpoint fields agree, so any difference in weight, hop count, or predecessor identifier settles the comparison.

We therefore only need to resolve a tie in all four fields. Simplicity of the graph identifies the last edge of the stored witness as the same edge $(u,v)$. Write $L_{\mathrm{old}}(u)$ for its historical prefix. Monotonicity gives $\lambda(u)\le L_{\mathrm{old}}(u)$.

Since the extension weights and hop counts agree, the prefix weights and hop counts agree as well. Appending the common last edge leaves only the comparison of the two reverse prefix sequences. We thus accept the tie: the fresh extension is no larger than the stored full label, without either historical sequence in storage.
\end{proof}

One of the four fields stores the hop count. To see why one word suffices, we bound the length of the path witnessing each accepted relaxation.

\begin{lemma}[Simple accepted witnesses]
\label{lem:simple_witnesses}
Starting from the source label and infinity elsewhere, every accepted relaxation has a simple historical path witness. Consequently every stored hop count is smaller than the number of vertices of the graph and fits in one metadata word.
\end{lemma}
\begin{proof}
We induct on accepted updates, starting from the source's simple witness. Suppose the current witness at $u$ is simple and extending it along $(u,v)$ repeats $v$.

We compare the candidate with the prefix ending at the earlier occurrence of $v$. Every label extends the label of its tail at the time of its relaxation, so that prefix was the shared label of $v$ when the relaxation creating it was accepted. Monotonicity places the current $\lambda(v)$ at or below that historical prefix. By contrast, the candidate adds a cycle of nonnegative weight and positive hop count, so it strictly exceeds the prefix even when the cycle has weight zero.

The acceptance test therefore rejects this candidate. Every accepted update preserves a simple historical witness, which closes the induction. A simple path has fewer edges than the graph has vertices, which bounds every stored hop count.
\end{proof}

\subsection{Capped feasibility and optimum-preserving tightening}
\label{sec:capped_feasibility}

A local search starts from a root whose label may already encode a path from the original source. The labels present at the start of the search limit which extensions of that path it may accept.

\begin{definition}[Capped feasibility]
\label{def:capped_feasibility}
Fix a local root $x$, a seed label $a:=\lambda(x)$, and a cutoff $B>\rho(a)$. Freeze, for analysis, the phase-start caps $c(v):=\lambda(v)$. A suffix starting at $x$, prefixed by the fixed witness of $a$, is feasible if every arrival on the new local suffix is at most its endpoint cap and has scheduling key below $B$. The inherited history before $x$ is not replayed against the new caps. Let $R$ be the feasible reachable set and let $D(v)$ be its minimum feasible full arrival label, with $D(v):=\infty$ outside $R$. Caps and seed histories are assumed coherent with the monotone relaxation history above.

For a scanned edge $(u,v)$, form $b$ by extending the current $\lambda(u)$ and accept precisely when
\[
\rho(b)<B\qquad\text{and}\qquad b\le\lambda(v).
\]
On acceptance, set $\lambda(v):=b$ and admit $v$.
\end{definition}

A cap constrains each prefix of a feasible path. To characterize an optimal path, we first ask whether improving a prefix can make the remaining suffix infeasible.

\begin{lemma}[Optimal feasible prefixes]
\label{lem:optimal_feasible_prefixes}
For the capped problem of Definition~\ref{def:capped_feasibility}, the following hold.
\begin{itemize}
    \item[(a)] Replacing a feasible prefix by a smaller feasible prefix preserves feasibility along an unchanged suffix.
    \item[(b)] Every canonical minimum feasible path has canonical minimum feasible prefixes.
\end{itemize}
\end{lemma}
\begin{proof}

{\bf Proof of Part (a).}
We first compare two prefixes after appending the same suffix. This adds the same weight and hop count to both labels. If those fields tie, the reversed suffix supplies the same leading identifiers, after which the comparison reaches the two prefixes. Their order is therefore unchanged.

Replacing a prefix by a smaller feasible prefix makes every subsequent arrival no larger. Each arrival still respects its endpoint cap and the cutoff, so the entire replacement remains feasible.

{\bf Proof of Part (b).}
To obtain an optimum, we restrict to simple suffixes. Removing a cycle does not increase weight and strictly decreases hop count, and there are only finitely many simple suffixes.

Now suppose an optimum contains a nonminimum feasible prefix. We replace that prefix by a smaller one. Feasibility survives by the preceding argument, while the label at the whole path's endpoint strictly decreases, a contradiction. The full tie order selects one optimum and hence consistent canonical prefixes.
\end{proof}

Feasibility depends on the labels at the start of a phase, but relaxations continually update those labels. Distinguishing the first admission of a vertex from its later updates lets us use one array for both purposes.

\begin{lemma}[Shared-array simulation]
\label{lem:shared_array_simulation}
During one local phase, suppose only $x$ is initially admitted, adjacency lists are scanned only at admitted vertices, and the outer caller performs no unrelated relaxations. The relaxation rule of Definition~\ref{def:capped_feasibility} implements an immutable phase-start cap test together with ordinary local tentative-label improvement.
\end{lemma}
\begin{proof}
We interpret the shared-label test in Definition~\ref{def:capped_feasibility} on the two sides of first admission:
\begin{itemize}
\item Before admission, no local update touches $v$, so $\lambda(v)=c(v)$. The acceptance test checks the frozen phase-start cap.
\item After admission, $\lambda(v)$ is the least label generated locally at $v$ so far, and $\lambda(v)\le c(v)$. The same test checks improvement of that local label and also enforces the cap.
\end{itemize}
We scan only admitted tails. Starting from the root witness and applying the appropriate case at each accepted extension therefore gives a feasible witness for every admitted vertex. An unadmitted finite cap only bounds that extension. It does not create another source.
\end{proof}

An accepted relaxation may lower the label at its endpoint. Before using that label as a cap in a later search, we need to know whether the tighter cap preserves the feasible optimum.

\begin{lemma}[Optimum-preserving cap tightening]
\label{lem:cap_tightening}
Let $h(v)$ be a generated feasible arrival label, or infinity if no such label is supplied. Tighten the caps by setting
\begin{equation*}
c'(v):=\min\{c(v),h(v)\}.
\end{equation*}
\begin{itemize}
    \item[(a)] The reachable set $R$ is preserved.
    \item[(b)] Every optimum label $D(v)$ is preserved.
\end{itemize}
\end{lemma}
\begin{proof}

{\bf Proof of Part (a).}
Tightening does not add feasible paths. For every $v\in R$, however, $D(v)\le c(v)$ and $D(v)\le h(v)$, because each finite $h(v)$ is a feasible witness. Therefore $D(v)\le c'(v)$.

By Lemma~\ref{lem:optimal_feasible_prefixes}-(b), every prefix of an old canonical optimum path has exactly the corresponding label $D(u)$. All of these prefixes satisfy the tightened caps, and their cutoff is unchanged. Every old optimum remains feasible, so every vertex of $R$ remains reachable. Tightening cannot make a new vertex reachable.

{\bf Proof of Part (b).}
For each $v\in R$, its old optimum is still feasible by the proof of Part (a), and tightening cannot introduce a smaller feasible label. Thus its optimum label remains $D(v)$.
\end{proof}

When a search outgrows its allotted capacity, the next phase starts again at the same root. Cap tightening lets us discard the old queue while keeping the label improvements it produced.

\begin{lemma}[Restart without label rollback]
\label{cor:restart_without_rollback}
Reset the scratch queues and local admission marks while retaining the shared label improvements and using the current labels as the next phase's caps.
\begin{itemize}
    \item[(a)] The reachable set and optimum labels are preserved across restarts.
    \item[(b)] The root seed is preserved, so every restarted phase solves the same rooted feasible problem.
\end{itemize}
\end{lemma}
\begin{proof}

{\bf Proof of Part (a).}
We apply Lemma~\ref{lem:cap_tightening}-(a)--(b) with $h$ equal to the best feasible labels generated by the preceding phase. Thus adopting the current shared labels as the next caps preserves the reachable set and optimum array.

{\bf Proof of Part (b).}
Any nonempty suffix returning to the root adds nonnegative weight and positive hop count, so it fails to improve that seed. The next phase therefore solves the same rooted problem. Repeating the argument covers any finite number of restarts.
\end{proof}

\subsection{Frontier certificates and cumulative discovery trees}
\label{sec:frontier_certificates}

A recursive call starts from several candidate roots, some of which may still have tentative labels. The canonical paths through these roots determine which vertices the call must complete.

\begin{definition}[Frontier targets and pivot attempts]
\label{def:frontier_targets}
Fix an enclosing engine's optimum $D$. A vertex is complete if its effective tentative label equals $D(v)$. An unadmitted vertex has effective label infinity. Let $T(B,S)$ consist of vertices with true scheduling key below $B$ whose canonical optimum suffix visits $S$. The local suffix, rather than the seed's pre-root history, defines visitation. Assume that every source in $S$ is admitted with current key below $B$, every target vertex has a complete canonical ancestor in $S$, and every vertex outside $T(B,S)$ with true key below $B$ is already complete. The last condition is the \emph{exterior-completeness invariant}.

A pivot attempt rooted at $x\in S$ uses seed $\lambda(x)$, cutoff $B$, and the relaxation rule of Definition~\ref{def:capped_feasibility}. It maintains a cumulative discovery set $C:=\{x\}$ and a set $E_C:=\varnothing$ of first-discovery edges. On the first admission of a new vertex to $C$, record that vertex and the edge admitting it before testing the stopping conditions. The attempt stops at the first hit of the existing forest, at $|C|=k$, or at local exhaustion, in that order of priority. The existing forest stays fixed during the entire attempt. The sets $C,E_C$ and shared label improvements survive any capacity restart. The attempt returns $(C,E_C,\sigma)$, where $\sigma\in\{\mathsf{forest},\mathsf{size},\mathsf{exhausted}\}$ records the stopping reason. Roots already in the existing forest need no attempt. For $k\ge2$, the pivot construction requires an attempt with this interface that computes minimum feasible labels throughout its local region on exhaustion.

For an attempt procedure $\mathcal A$ satisfying this interface, Algorithm~\ref{alg:pivot_forest} displays the pivot construction $\textsc{PivotForest}_{\mathcal A}(B,S,k;\lambda)$, which calls $\mathcal A(x,B,k,F;\lambda)$ at each root. The forest and accounting arguments apply to every such $\mathcal A$.
\end{definition}

\begin{algorithm}
\caption{Pivot and forest construction}
\label{alg:pivot_forest}
\begin{algorithmic}[1]
\Procedure{PivotForest$_{\mathcal A}$}{$B,S,k;\lambda$} \Comment{Lemma~\ref{lem:cumulative_forest}}
\State $F \gets \varnothing$; $Q \gets \varnothing$; $W \gets \varnothing$
\For{each $x\in S$ in the supplied order}
    \If{$x\notin V(F)$}
        \State $(C,E_C,\sigma) \gets \mathcal A(x,B,k,F;\lambda)$.
        \If{$\sigma=\mathsf{exhausted}$}
            \State $Q \gets Q\cup\{x\}$; $W \gets W\cup C$
        \Else
            \State $V(F) \gets V(F)\cup C$; $E(F) \gets E(F)\cup E_C$
        \EndIf
    \EndIf
\EndFor
\State $P \gets S\setminus Q$
\State Partition $F$ into supporting trees $F_1,\ldots,F_p$ and assign each $x\in P$ a containing-tree index $j(x)$. Use $p \gets 0$ if $F=\varnothing$.
\State Initialize $P_j \gets \varnothing$ for $1\le j\le p$.
\For{each $x\in P$}
    \State $j \gets j(x)$; $P_j \gets P_j\cup\{x\}$
\EndFor
\State \Return $(\{(P_j,F_j)\}_{j=1}^{p},Q,W)$
\EndProcedure
\end{algorithmic}
\end{algorithm}

A complete vertex also gives us information about the path by which it was reached. Tracing its canonical suffix back to the local root identifies earlier vertices whose labels are already exact.

\begin{lemma}[Hereditary completeness]
\label{lem:hereditary_completeness}
In an enclosing environment obeying the admission discipline of Lemma~\ref{lem:shared_array_simulation}, let an admitted vertex $v$ have finite label $D(v)$. Every vertex on the canonical local optimum suffix from that environment's root to $v$ is admitted and complete. This statement concerns the local suffix, not the seed's pre-root history.
\end{lemma}
\begin{proof}
We trace the generation witness of $D(v)$ back to this environment's root. Only the root has initial admission, and each update extends an admitted tail. Hence every vertex on the witness's local suffix receives admission. This includes the case where $D(v)$ was already an inherited cap and $v$ was admitted by an equality relaxation. The accepted candidate is at most the cap and, being feasible, at least $D(v)$, so it equals $D(v)$ and extends the label of an admitted tail.

A full label determines its path, so this witness is the canonical optimum suffix. Lemma~\ref{lem:optimal_feasible_prefixes}-(b) identifies each generated prefix at $u$ with $D(u)$.

After admission, we have both bounds on the label at $u$: monotonicity keeps it at most $D(u)$, while feasibility keeps it at least $D(u)$. Thus $u$ remains complete. Notification records the update in every affected enclosing environment, so the same argument applies to each environment's own local suffix.
\end{proof}

A pivot attempt follows feasible paths, whereas membership in the target set depends on canonical paths. To relate the two, we use the completeness of vertices outside the target set whose true keys lie below the cutoff.

\begin{lemma}[Target containment]
\label{lem:target_containment}
Under the frontier hypotheses of Definition~\ref{def:frontier_targets}, every vertex admitted by a local capped attempt rooted at $x\in S$ lies in $T(B,S)$. The same holds for the head of every accepted relaxation below $B$ whose tail is admitted and lies in $T(B,S)$, in particular for the level-zero scans, the boundary scans, and the final scan of $W'$ in Algorithm~\ref{alg:bmssp}.
\end{lemma}
\begin{proof}
We take an accepted candidate and suppose its endpoint lies outside $T(B,S)$. The candidate is a feasible enclosing-engine path below $B$, so the endpoint's true key is also below $B$.

Exterior completeness then makes the current endpoint label optimal. Acceptance forces the candidate to equal that optimum, including the full tie order. Yet its path visits the attempt root $x\in S$, which places the endpoint in $T(B,S)$. This contradiction proves containment.

For an accepted relaxation with an admitted tail $u\in T(B,S)$, the same comparison shows that the candidate equals the optimum of its endpoint. By Lemma~\ref{lem:optimal_feasible_prefixes}-(b), its prefix at $u$ is the canonical optimum of $u$, whose suffix visits $S$. The endpoint therefore lies in $T(B,S)$ as well.
\end{proof}

When a search from a complete root exhausts its queue, the algorithm discards the root. To justify this step, we identify which of its canonical descendants the search has already completed.

\begin{lemma}[Exhausted-root certificate]
\label{lem:exhausted_root_certificate}
Suppose a local attempt rooted at a complete $x\in S$ exhausts its feasible region.
\begin{itemize}
    \item[(a)] It includes and completes every canonical enclosing-optimum descendant of $x$ below $B$.
    \item[(b)] Consequently discarding exhausted roots into a set $Q$ and adding their regions to $W$ leaves a frontier consisting of completed vertices in $W$ and retained roots in $S\setminus Q$.
\end{itemize}
\end{lemma}
\begin{proof}

{\bf Proof of Part (a).}
We fix a complete root $x$ and follow an enclosing canonical optimum through it. At the start of the attempt, every suffix arrival satisfies $D(v)\le\lambda(v)=c(v)$ and lies below $B$. Thus this suffix is locally feasible.

Because the seed at $x$ is exact, no local feasible path improves on the enclosing optimum $D$. Exhaustion therefore reaches and completes every such descendant. Lemma~\ref{cor:restart_without_rollback}-(a)--(b) preserves this conclusion across capacity restarts.

{\bf Proof of Part (b).}
For each target vertex, choose its original complete witnessing source:
\begin{itemize}
\item If the source remains in $S\setminus Q$, it remains a frontier witness.
\item If the source enters $Q$, its exhausting search includes the target vertex in $W$ and completes it.
\end{itemize}
Only exhaustion from a complete witness justifies the second case. A size limit, timer, or forest hit supplies no such certificate.

We do not infer completeness of every vertex in $W$ from this argument. An exhausted attempt from an incomplete root may leave stale enclosing labels. Lemma~\ref{thm:capped_return_contract}-(a) proves completeness of the returned subset $W'$ using the exact-prefix and exterior-completeness invariants.
\end{proof}

A pivot attempt can also reach its size limit or meet a tree built by an earlier attempt. Its first-discovery edges let us count the work and group the roots that the algorithm retains.

\begin{lemma}[Cumulative forest accounting]
\label{lem:cumulative_forest}
In procedure \textnormal{\textsc{PivotForest}} (Algorithm~\ref{alg:pivot_forest}), the following hold.
\begin{itemize}
    \item[(a)] The first-discovery edges form a tree on $C$. At the first forest hit this tree intersects the existing forest in exactly the hit vertex, so their union is a weak tree.
    \item[(b)] If an attempt of cumulative size $r$ costs $O(rf(r))$, for nondecreasing $f\ge1$, all pivot attempts cost
\[
O(|F|f(k)+kf(k)|Q|+|S|),
\]
where $F$ is the final forest and $Q$ is the discarded-root set.
    \item[(c)] The retained forest can also be partitioned in $O(|F|)$ time into $O(\frac{|F|}{k})$ connected supporting trees of sizes in $[k,3k)$, with disjoint edges and total vertex occurrences $O(|F|)$. Each retained root can be assigned to exactly one containing tree in the same time bound.
\end{itemize}
\end{lemma}
\begin{proof}

{\bf Proof of Part (a).}
Each first-discovery edge attaches a new vertex to an admitted tail, so these edges form a tree on $C$. At the first forest hit, this tree meets the old forest only at the hit vertex, so their union is again a weak tree. Keeping $C$ and its discovery edges across restarts preserves this fact even if a restart changes the exploration order.

{\bf Proof of Part (b).}
We charge the attempts according to their stopping reason:
\begin{itemize}
\item A size-limited attempt contributes $k$ new forest vertices.
\item A merge attempt of size $r\ge2$ contributes $r-1$ new vertices. Its size satisfies $r\le2(r-1)$.
\item An exhausted attempt has size less than $k$ and charges its work to its discarded root in $Q$.
\end{itemize}
Consequently the total cumulative size of retained and merged attempts is at most $2|F|$. Monotonicity of $f$, together with the initial processing of $S$, gives the pivot bound in Lemma~\ref{lem:cumulative_forest}-(b).

We count new vertices relative to the retained forest, without subtracting $W$: equality relaxations may re-admit vertices from an earlier exhausted region, so $F$ and $W$ need not be disjoint.

{\bf Proof of Part (c).}
We next partition the forest into supporting groups using the construction of~\cite{dmsy26}. To expose its degree-independent cost, root each weak tree and process it in postorder. At a vertex, concatenate the residual pieces from its children with the current vertex as their common attachment. Each child residual has fewer than $k$ vertices.

Whenever the accumulated piece first reaches $k$ vertices, we report it and restart with the current vertex alone. Each reported piece has fewer than $2k$ vertices.

At the root, we merge the final residual into the last reported piece. No piece is reported after that one, so the vertex at which it was reported stays in every later residual up to the root. The final residual has fewer than $k$ vertices and contains that attachment vertex, so the merge stays connected and has fewer than $3k$ vertices. Every retained component has at least $k$ vertices, which guarantees a reported piece to receive the residual.

Let $p$ be the number of pieces.
We have
\begin{equation}
p(k-1)\le |E(F)|,\qquad \sum_{j=1}^p |V(F_j)|=|E(F)|+p.
\label{eq:forest_piece_counts}
\end{equation}
The inequality follows because each piece contains at least $k-1$ edges and the pieces have disjoint edge sets. The equality follows by summing $|V(F_j)|=|E(F_j)|+1$ over the edge partition of $F$.

By Eq.~\eqref{eq:forest_piece_counts}, $p=O(\frac{|F|}{k})$ and the total vertex material is $O(|F|)$, independently of degree.

For the implementation, we concatenate linked residual lists in constant time using their stored sizes. We materialize each reported piece once, then traverse the pieces and assign each still-unassigned retained root to a containing tree through its local handle. This takes $O(|F|)$ time. Empty root groups require no queue representative, but their forest material still pays for partition construction.
\end{proof}

\subsection{The recursive capped engine}
\label{sec:capped_engine}

The recursive search keeps one queue representative for each group of roots. At each level it passes a block of low-key vertices to the level below, together with a cutoff for that block.

\begin{definition}[Recursive capped engine]
\label{def:capped_engine}
Choose integers $t\ge512$ and $2\le k\le \frac{t}{256}$, so in particular $24k\le t$. At level $\ell$ use
\[
|S|\le t^2 2^{\ell t},\qquad
\Lambda_\ell:=t^3 2^{\ell t},\qquad
M_\ell:=8t2^{(\ell-1)t}\quad(\ell>0).
\]
For a pivot attempt $\mathcal A$, procedure $\textsc{BMSSP}_{\mathcal A}(B,S,\ell,t,k;\lambda)$ in Algorithm~\ref{alg:bmssp} receives $t,k$ explicitly and calls $\textsc{PivotForest}_{\mathcal A}$. Its recursive calls retain $t,k$ and the same choice of $\mathcal A$, and decrease only the level $\ell$. The return and accounting arguments below apply to every such $\mathcal A$. All relaxations use Definition~\ref{def:capped_feasibility}, source occurrences and queue records carry the tokens of Definition~\ref{def:provenance_tokens}, and nonbase queues use Definition~\ref{def:blocked_queue}.

At level zero, run bounded Dijkstra from $S$, stopping after more than $t^3$ settled vertices or queue exhaustion. Its queue is keyed by the current shared labels of the sources at call entry, not by their pulled stored keys. Relax every outgoing edge of a settled vertex below $B$ before testing the stopping condition. Return $(b,U,\mathcal D)$, where $U$ is the settled set, $\mathcal D$ is the residual queue, and $b$ is its minimum remaining queue key, or $B$ if it is empty.

The recursion starts at $L:=\lceil\frac{\beta(Q)}{t}\rceil$, where $\beta(Q):=\max\{4,\lceil\lg Q\rceil\}$, using a supplied size upper bound $Q$ on the local instance. The degree bound, parameters, top level, and original identifier convention form a frozen parameter header. They are not recomputed from the vertices visited by a phase.
\end{definition}

\begin{algorithm}
\caption{The recursive capped engine}
\label{alg:bmssp}
\begin{algorithmic}[1]
\Procedure{BMSSP$_{\mathcal A}$}{$B,S,\ell,t,k;\lambda$} \Comment{Lemma~\ref{thm:capped_return_contract}}
\If{$\ell=0$}
    \State Run the level-zero search of Definition~\ref{def:capped_engine}. \Return its triple $(b,U,\mathcal D)$.
\EndIf
\State $(\{(P_j,F_j)\}_{j=1}^{p},Q,W) \gets \Call{PivotForest$_{\mathcal A}$}{B,S,k;\lambda}$ \Comment{Algorithm~\ref{alg:pivot_forest}}
\State Initialize a blocked queue $\mathcal D$ with parameter $M_\ell$; $U\gets\varnothing$.
\For{each nonempty group $P_j$}
    \State Choose a minimum-current-key $p_j\in P_j$; $\alpha_j\gets\rho(\lambda(p_j))$. Insert $(p_j,\alpha_j)$ into $\mathcal D$.
\EndFor
\State $b\gets$ the minimum stored key in $\mathcal D$, or $B$ if empty.
\While{$\mathcal D\ne\varnothing$ and $|U|\le\Lambda_\ell$}
    \State Pull the smallest block $K$ from $\mathcal D$ with separating cutoff $B_i$.
    \State $S_i\gets K$, keeping every pulled record and its token. Use endpoint stamps.
    \State For each current representative in $K$, add its group members with current key below $B_i$ to $S_i$, keeping their input tokens and omitting endpoints already present.
    \State $(b_i,U_i,\mathcal D_i)\gets\Call{BMSSP$_{\mathcal A}$}{B_i,S_i,\ell-1,t,k;\lambda}$ \Comment{Algorithm~\ref{alg:bmssp}}
    \State Delete all vertices of $U_i$ from the parent queue $\mathcal D$, if present. Merge $\mathcal D_i$ into $\mathcal D$.
    \State Mark exactly the groups whose representatives lie in $U_i$. Remove $U_i$ from all groups.
    \For{each outgoing edge $(u,v)$ with $u\in U_i$}
        \State Form its candidate label $\eta$ from the current label of $u$.
        \If{$B_i\le\rho(\eta)<B$ and the shared relaxation succeeds}
            \State Insert $(v,\rho(\lambda(v)))$ into $\mathcal D$.
            \State If $v\in P_j$ for an unmarked group and $\rho(\lambda(v))<\alpha_j$, set $p_j\gets v$ and $\alpha_j\gets\rho(\lambda(v))$.
        \EndIf
    \EndFor
    \State Rescan each marked nonempty $P_j$, choose a minimum-current-key $p_j$, set $\alpha_j\gets\rho(\lambda(p_j))$, and insert $(p_j,\alpha_j)$ into $\mathcal D$.
    \State $U\gets U\cup U_i$; $b\gets b_i$.
\EndWhile
\State If $\mathcal D=\varnothing$, set $b\gets B$.
\State Insert every original source whose current key lies in $[b,B)$ into $\mathcal D$, using its current key and input token.
\State $W'\gets\{v\in W\setminus U:\rho(\lambda(v))<b\}$.
\State Relax outgoing edges of $W'$ below $B$. Insert accepted heads with keys at least $b$ into $\mathcal D$.
\State \Return $(b,U\cup W',\mathcal D)$.
\EndProcedure
\end{algorithmic}
\end{algorithm}

\paragraph{Saved representative thresholds.}
A child call can lower a label while its parent still stores the old queue key. A complete canonical ancestor whose record has the true key already serves as a frontier witness. For group members without such a witness, the saved threshold ensures that their representative is pulled in time.

\begin{definition}[Saved-threshold invariant]
\label{def:saved_thresholds}
A \emph{pending} or \emph{unresolved} target is a target that has not been returned, even if its shared label is already exact. All frontier assertions below cover these complete-but-unreturned vertices as well.

A queue record is an \emph{accurate complete witness} if its vertex is complete and its stored scheduling key equals the true scheduling key. Let $A(\mathcal D)$ be these vertices and define the proof-only set
\[
\widehat P_j:=\{v\in P_j:\text{the canonical optimum path to $v$ misses }A(\mathcal D)\}.
\]
The saved-threshold invariant is
\[
\operatorname{storedkey}(p_j)\le\alpha_j
\le\min_{v\in\widehat P_j}\rho(\lambda(v)).
\]
The minimum over an empty set is infinity. Each remaining group has a representative in the queue. In addition, unresolved target vertices have a complete canonical ancestor in $A(\mathcal D)\cup\bigcup_j P_j$, residual and group keys are at least the progress bound, and the exterior-completeness invariant holds.

\end{definition}

The simultaneous proof of these invariants is given in Lemma~\ref{thm:capped_return_contract}-(a).

\subsection{Inherited histories and nested environments}
\label{sec:inherited_histories}

A local search may run on a subgraph even though its initial labels came from paths in the ambient graph. We specify which parts of this history the search inherits.

\begin{definition}[Admissible capped instances]
\label{def:admissible_instances}
Fix a simple ambient graph $H$ and a monotone history of valid relaxations from its original source. A local instance consists of a subgraph $G$ of $H$, a root, its current seed label, a cutoff, and the current shared labels as caps. Original vertex identifiers are retained. Historical witnesses belong to $H$ and need not belong to $G$. Subsequent local extensions use only edges of $G$. The seed and caps are supplied labels, not paths that the local algorithm must construct anew. The computational representation is still the four fields of Definition~\ref{def:path_labels}. Full inherited paths occur only in the correctness proof.
\end{definition}

A pivot search introduces a new root and a new set of caps. The enclosing searches still share the label array, so they must learn which vertices the new search admits.

\begin{definition}[Execution environments]
\label{def:execution_environment}
An \emph{environment} is one fixed root/cap problem, with its own admitted set and optimum array $D$. All recursive BMSSP calls in that environment refer to the same $D$. A pivot search starts a new environment, whereas descending one BMSSP level does not. Only the innermost active environment performs relaxations. Every successful relaxation first writes the shared label and records admission in all active environments and cumulative attempt monitors, and only then delivers an interruption signal. If several monitors request a stop, the outermost requesting monitor wins. Within one monitor, a forest hit or discovery limit takes precedence over a capacity restart. Suspended queues keep their stored keys unchanged. The interruption checkpoint is after all label writes, admission notifications, and cumulative-monitor updates, but before dispatch of any stop or the subsequent queue insertion. A local bound includes its descendants and all their notifications, but excludes execution of a passive ancestor's control flow after the local attempt stops. Destruction of an ancestor's preexisting arena is charged to that ancestor's earlier record creations.

Control frames, including suspended BMSSP frames, are explicit owned records.
Each environment stores a continuation for each of its stopping monitors and an
ownership list for its frames and scratch records. Creation or reinitialization
prepays a constant amount for later unlinking, handle clearing, and destruction.
This applies to control frames as well as queue and forest records. On a stop,
the winning monitor is selected during notification and dispatch jumps to its
saved continuation. Discarded frames are reclaimed from ownership lists, without
resuming their search code or deleting their queue records one by one through a
balanced tree. A wrapper's cumulative discovery records are owned separately
from a phase and survive a capacity restart. Shared labels are never rolled back.
In particular, BMSSP stack length is not charged to the notification scan.
Throughout the accounting below, a charged cost includes these deposits when
records are created. Their eventual reclamation spends the deposits and is not
charged a second time. Every prefix has nonnegative remaining credit. Cleanup of
preexisting external records uses the external owner's deposits.

The admissible instance class also permits passive shadow observers: they perform the same bounded direct-address membership and cumulative-discovery bookkeeping, with external stop delivery disabled. They never resume ancestor control code or scan a preexisting ancestor forest. This extension is used only for hypothetical complete executions in the interruption proof. The actual wrapper delivers stops at the specified checkpoint.

The local engine receives a fixed \emph{parameter header}: a capacity upper bound
$Q$, its total observer budget, its parameters $t,k$, its top BMSSP level, the advertised constant degree
bound, and the original identifier/addressing convention. It never scans the
ambient vertex set to recompute any of these quantities. In particular, it does
not recompute the graph's actual maximum degree during a restart or a restriction.
It inspects an adjacency list only at an admitted vertex. Queues, groups, lists,
and marks are created lazily and tracked in ownership/touched-record lists.
\end{definition}

A relaxation made by the innermost search also changes the labels seen by its callers. Its path must therefore respect the caps and cutoff of every enclosing search.

\begin{lemma}[Preservation in enclosing environments]
\label{lem:nested_feasibility}
For the execution environments of Definition~\ref{def:execution_environment}, suppose an inner environment starts from a vertex admitted in its enclosing environment, with its current shared label, current shared caps, and a cutoff no larger than the enclosing cutoff.
\begin{itemize}
    \item[(a)] Every accepted inner update is feasible in each enclosing environment.
    \item[(b)] Cap tightening and restarting preserve each enclosing optimum.
    \item[(c)] With at most $j$ nested environments, admission notification costs $O(j)$ per successful relaxation.
\end{itemize}
\end{lemma}
\begin{proof}

{\bf Proof of Part (a).}
We induct on accepted inner edges, starting from the root witness, which is feasible in each enclosing environment. For the next scanned edge $(u,v)$, induction and notification place $u$ in every enclosing admitted set with a feasible current label.

The inner acceptance guards enforce both enclosing constraints. The candidate is at most the shared head label, which is no larger than each frozen cap. Its key lies below the inner cutoff, which is no larger than each enclosing cutoff. Extending the witness at $u$ therefore proves feasibility in every enclosing environment.

Because notification records these admissions before interruption, the induction also covers an interrupted prefix.

{\bf Proof of Part (b).}
We then apply Lemma~\ref{lem:cap_tightening}-(b) separately in each environment to preserve its optimum through tightening and restarting.

{\bf Proof of Part (c).}
Each successful relaxation inspects one notification slot per active environment, at cost $O(j)$. Feasibility transfers outward, but the argument never identifies an inner optimum with an enclosing optimum.
\end{proof}

A search may stop after inspecting only a small part of the graph. To bound the work already done, we reproduce that part of the execution on a graph whose size depends on the number of admissions.

\begin{lemma}[Small completion with inherited histories]\label{lem:small_completion}
For admissible instances as in Definition~\ref{def:admissible_instances} with maximum out-degree $\Delta=O(1)$, an execution prefix of a phase with at most
$q+1$ admissions is the identical prefix of an admissible local instance with at
most $(\Delta+1)(q+1)$ vertices, using the same parameter header and initial
shared labels. Original identifiers and inherited history are preserved. The
statement includes suspended observer state. A stopping prefix ends after all
triggering notifications and before abort dispatch or subsequent queue insertion.
An ordinary prefix may end at an operation boundary.
\end{lemma}
\begin{proof}
We construct a comparison instance from the state that the prefix actually inspects. Let $A$ contain the admitted vertices at the end of the prefix, including the root.

We keep every outgoing edge of $A$ and every head of those edges, and give retained vertices outside $A$ empty outgoing lists. The resulting subgraph of $H$ has at most $(\Delta+1)|A|$ vertices.

To replay the prefix, we initialize with the phase-start labels. Using end-of-prefix labels instead exposes improvements before the operations that produce them. We also retain the header and original identifiers, together with the initial monitor fields, forest membership, and cumulative admissions.

Inherited witnesses stay in the ambient graph $H$. Since a predecessor field is an identifier word, its vertex need not become a local vertex.

By the parameter-header convention of Definition~\ref{def:execution_environment}, the engine reads the graph only through adjacency lists of admitted vertices. It reads labels, marks, queue records, and group records only at vertices of $A$ and at heads of edges leaving $A$. Every adjacency inspection has its tail in $A$ and sees the same list. Every referenced label or membership field also agrees. Induction over deterministic operations therefore gives the same updates, notifications, and queue actions throughout the prefix.

We retain the ambient history that supplies the seed and caps. Lemma~\ref{lem:four_field_test} tests their extensions from the stored fields, without reconstructing a witness inside the restricted graph. Thus the comparison instance remains admissible.

At a stopping prefix, both executions reach the same post-notification checkpoint with the same stop requests. For the hypothetical complete execution, disable subsequent external stop delivery. Shadow observers continue their bounded bookkeeping without resuming ancestor control or scanning a preexisting forest.

With empty outgoing lists outside $A$, we continue on a finite admissible instance and create observer fields only for vertices that receive admission. If the prefix ends during a relaxation, continuation starts after the same shared write and notifications. An ordinary operation-boundary prefix needs no different construction.

Finally, we preserve the cost comparison through ownership: local record creation prepays local destruction. Cleanup of a preexisting external arena spends its external owner's deposits and adds no new local charge.
\end{proof}

\subsection{A simultaneous correctness induction}
\label{sec:engine_correctness}

A recursive call can stop with some targets still pending. Its caller needs to know which vertices are complete and how to reach the pending targets from the remaining queue.

\begin{definition}[Capped return contract]
\label{def:return_contract}
Fix one environment's optimum $D$ and input $(B,S)$ satisfying Definition~\ref{def:frontier_targets}. The return contract states that $U=T(b,S)$ is complete, the residual queue represents unconsumed sources and valid boundary relaxations below $B$, and every residual current label and stored key is at least $b$. If $b<B$, it also requires $|U|>\Lambda_\ell\ge t|S|$, while always $|U|=O(\Lambda_\ell)$. In addition, the residual queue satisfies the following conditions: records belong to $T(B,S)\setminus U$. Every source in $S\setminus U$ has a record. Every unresolved target has a complete canonical ancestor whose record is stored at its true key. And
\[
b\le\rho(\lambda(v))\le\operatorname{storedkey}(v)<B
\]
for every residual record. For each unconsumed original source, its residual stored key is also at most its current key at call entry. This entry-key bound is part of the inductive contract. A record stored above its current key is allowed. Relaxation and queue insertion are separate operations. Hidden label changes do not decrease a stored key.
\end{definition}

The cutoff returned by a child determines how far its parent can advance. We prove exactness below that cutoff together with the queue invariant needed for the next child.

\begin{lemma}[Capped return contract]
\label{thm:capped_return_contract}
With the admission discipline of Lemma~\ref{lem:nested_feasibility}-(a), procedure \textnormal{\textsc{BMSSP}} (Algorithm~\ref{alg:bmssp}) with the parameters of Definition~\ref{def:capped_engine} has the following guarantees at every level, as required by Definition~\ref{def:return_contract}.
\begin{itemize}
    \item[(a)] It returns precisely the complete set $T(b,S)$ and satisfies the residual-queue conditions of Definition~\ref{def:return_contract}.
    \item[(b)] If $b<B$, then $|U|>\Lambda_\ell\ge t|S|$, and always $|U|\le 3\Lambda_\ell$ for $t\ge512$.
\end{itemize}
This correctness statement only requires correct exhaustive local searches and the specified stopping rules. It is independent of their running time.
\end{lemma}
\begin{proof}

{\bf Proof of Part (a).}
We induct on the BMSSP level, carrying a complete canonical witness for each unresolved target. We keep this witness even when the target already has its exact label but does not yet belong to the return set. Write $\kappa(v):=\rho(D(v))$.

At level zero, we follow a canonical path from a complete input source to an unresolved target, and let $h$ be its first unsettled vertex. If $h$ is the source itself, it is in the queue with its true key. Otherwise its predecessor $u$ on the path is settled and complete, and the scan of $u$ formed the candidate $D(h)$ along $(u,h)$. This candidate is at most the current $\lambda(h)$, so the relaxation rule of Definition~\ref{def:capped_feasibility} accepts it, also when $\lambda(h)$ already equals $D(h)$ through a hidden update or an inherited cap. The scan therefore puts $h$ in the queue with its true key, which is no larger than the target's true key.

Suppose the minimum queue vertex $z$ is not complete. Its true key is at most its current key and hence below $B$, so the exterior-completeness invariant of Definition~\ref{def:frontier_targets} places $z$ in $T(B,S)$. Thus $z$ is itself an unresolved target. Let $h$ be the first unsettled vertex on its canonical path, as in the preceding paragraph. If $h=z$, that paragraph shows that $z$ already carries its optimum label, contradicting incompleteness. Otherwise $h$ is a proper canonical ancestor of $z$ in the queue, and its key is strictly smaller than the true key of $z$, which is at most the current key of $z$. This contradicts the minimality of $z$. Hence the minimum queue vertex is complete.

Applying this argument after each settled vertex, and relaxing all its outgoing edges before stopping, gives exactness and the residual frontier. Equal relaxations preserve this argument for zero weights and inherited caps. An unadmitted finite cap never acts as a source. The level-zero workload bound follows from the stopping rule.

\paragraph{Step 1: Establish the Retained Frontier.}
For a nonbase call, we freeze $P^0:=S\setminus Q$ after pivot construction and set $T_P:=T(B,P^0)$. Lemma~\ref{lem:exhausted_root_certificate}-(a) completes every relevant descendant of a discarded complete source. Together with the input exterior-completeness invariant, this makes every vertex outside $T_P$ below $B$ complete.

To find a retained witness, we fix $v\in T_P$ and its original complete witnessing source $q\in S$. If $q\in P^0$, it still witnesses $v$. Otherwise choose a retained root $p\in P^0$ on the canonical suffix to $v$. Both roots lie on that suffix, and their order gives two cases:
\begin{itemize}
\item If $q$ precedes $p$, exhaustion from $q$ completes $p$ by Lemma~\ref{lem:exhausted_root_certificate}-(a).
\item If $p$ precedes $q$, hereditary completeness of the admitted complete vertex $q$ completes $p$ by Lemma~\ref{lem:hereditary_completeness}.
\end{itemize}
Thus every vertex of $T_P$ has a complete retained ancestor. No claim about other discarded roots enters this argument.

We maintain three assertions at each loop boundary, always with this same $T_P$:
\begin{enumerate}
\item $U=\{v\in T_P:\kappa(v)<b\}$ and every vertex of $U$ is complete. Remaining group and queue vertices lie in $T_P\setminus U$, so their true keys are at least $b$.
\item Every unresolved vertex of $T_P$ has a complete canonical ancestor among the remaining groups and the accurate queue witnesses $A(\mathcal D)$.
\item Each nonempty group has a representative record and satisfies Definition~\ref{def:saved_thresholds}. Current and stored keys are at least $b$, and each current key is at most its stored key.
\end{enumerate}
Initially each representative is a current group minimum. A complete group ancestor of any target has key at least the minimum representative key. Thus no target has true key below the initial $b$, as required for $U=\varnothing$. The retained-ancestor argument gives the second assertion, and choosing group minima gives the third.

\paragraph{Step 2: Verify the Child Input.}
Let $B_i$ be the separator of the next pull, and take an unresolved target with true key below $B_i$. Its complete ancestor from the second invariant reaches the child in one of two ways:
\begin{itemize}
\item If its canonical path meets an accurate queue record, that record is stored at its true key, which is at most the target's true key and hence below $B_i$. The pulled block contains every record with stored key below its separator $B_i$, so it contains this record.
\item Otherwise its witnessing group root is necessary in Definition~\ref{def:saved_thresholds}. The saved threshold puts its representative below $B_i$. Pulling that representative expands the group and includes the complete root.
\end{itemize}
In both cases $S_i$ contains a complete witness. We now check the child source budget.
We have
\begin{equation}
|S_i| \le (1+3k)M_\ell = 8t(1+3k)2^{(\ell-1)t} \le t^2 2^{(\ell-1)t}.
\label{eq:child_source_size}
\end{equation}
The first step follows from counting at most $M_\ell$ pulled records, each contributing its own vertex and at most $3k$ group members. A group has fewer than $3k$ roots because its supporting tree has fewer than $3k$ vertices by Lemma~\ref{lem:cumulative_forest}-(c).
The second step follows from $M_\ell=8t2^{(\ell-1)t}$ from Definition~\ref{def:capped_engine}.
The last step follows from $8(1+3k)\le t$, since $k\le t/256$ and $t\ge512$.

Every source in $S_i$ is admitted, lies in $T_P\setminus U$, and has current key below $B_i$. Its canonical descendants stay in $T_P$. Their true keys increase along the suffix, so none belongs to $U$.

We also need exterior completeness for the child. The preceding case split places every remaining target below $B_i$ in the child target. A vertex below $B_i$ outside that target therefore lies either outside $T_P$ or in $U$, and both regions are complete. This supplies the remaining child input hypothesis. Lemmas~\ref{lem:nested_feasibility}-(a) and~\ref{lem:target_containment} supply target containment for its nested pivot attempts.

\paragraph{Step 3: Restore the Frontier and Saved Thresholds.}
With the source-size bound in Eq.~\eqref{eq:child_source_size} and exterior completeness, we apply the child contract to return exactly the remaining vertices of $T_P$ below $b_i$. Since child sources and their canonical descendants have true key at least $b$, we also have $b_i\ge b$.

We delete all of $U_i$ from the parent queue, including unpulled records. The enlarged $U$ is now the exact prefix below $b_i$, and every surviving group or record has true key at least $b_i$.

We replace a consumed complete witness by following its canonical path toward the unresolved vertex. The first head outside $U_i$ determines where its new witness comes from:
\begin{itemize}
\item Below $B_i$, the head is an unresolved target of the child, so the child's residual contract supplies an accurate record at one of its complete canonical ancestors. The canonical path to the head is a prefix of the canonical path to the unresolved vertex, so this record is also a witness for that vertex.
\item In $[B_i,B)$, the parent's explicit boundary scan inserts an accurate record.
\end{itemize}
In each case, we use a tight candidate, which is feasible and at most the shared head label. Equality admits the head if it lacks admission. No feasible update improves on $D$, so an accurate record stays accurate until consumption.

This split concerns the first exiting head, not the final target. An unresolved endpoint above $B_i$ may still need the child's residual witness below $B_i$.

To restore the saved thresholds, we ask which surviving roots still need their group representative. A root loses that need in either of two ways:
\begin{itemize}
\item An accurate record already covers it before the child call. Frontier replacement preserves that coverage unless the child consumes the root.
\item The child changes its label. Lemma~\ref{lem:target_containment} puts the root in the child target, so the child either consumes it or leaves an accurate residual ancestor.
\end{itemize}
A root that remains necessary therefore receives no hidden child decrease. If the parent also leaves its label unchanged, the old bound $\alpha_j\le\rho(\lambda(v))$ persists.

For an explicit parent boundary decrease, if the new key falls below $\alpha_j$, the parent sets $\alpha_j$ to that key and makes this root the representative. The same boundary operation inserts its accurate record. Otherwise the old threshold still bounds the new key. Thus each necessary root satisfies the upper bound on $\alpha_j$ in Definition~\ref{def:saved_thresholds}.

We complete the threshold check by tracing the representative record itself:
\begin{itemize}
\item If the child consumes it, a rescan selects a new current minimum.
\item If the pull takes it but the child does not consume it, source reinsertion returns a record with key at most its old stored key.
\item If the pull leaves it in the parent, the record persists.
\end{itemize}
These cases establish representative presence and both saved-threshold inequalities, completing the loop induction.

\paragraph{Step 4: Verify the Return Set and Residual Queue.}
If the queue empties, no group remains. The maintained frontier then exhausts $T_P$, so $b:=B$ is valid.

To verify the final contribution from $W$, we take a vertex of $W\setminus U$ with current key below $b$. Its true key is also below $b$. If it lies in $T_P$, exact-prefix completeness already places it in $U$. Otherwise exterior completeness makes it complete. Hence every vertex of $W'$ is complete.

Conversely, a vertex of $T(B,S)\setminus T_P$ below $b$ has a discarded complete source witness. The exhausted-root certificate places it in $W'$, unless it already belongs to $U$. The final return set is therefore precisely $T(b,S)$.

We reinsert every unconsumed original source and recover the final frontier from each target's original complete source witness. If that source is unconsumed, its reinserted record is accurate. If it is consumed, follow the canonical path to the first unconsumed vertex. The loop frontier supplies an accurate record after retained-target material, and the final scan of $W'$ supplies one after discarded-search material.

With an exact-prefix return, we have current key at least $b$ on every residual record. Its stored key dominates its current key, while the insertion test keeps the stored key below $B$.

Reinserting an original source uses its current shared key, which never increases from call entry. A successful final-scan update inserts its new current key, and duplicate resolution keeps the smaller stored key. This proves the entry-key bound and the residual contract.

{\bf Proof of Part (b).}
We now bound the return size and finish the termination argument.

\paragraph{Step 5: Bound the Workload and Prove Termination.}
A partial nonbase return requires the loop to exceed $\Lambda_\ell$. The last child adds at most $3\Lambda_{\ell-1}$ vertices, and $|W|\le k|S|\le\frac{\Lambda_\ell}{24}$.
We have
\[
|U|\le\Lambda_\ell+3\Lambda_{\ell-1}+\Lambda_\ell/24=(1+3\cdot2^{-t}+1/24)\Lambda_\ell\le3\Lambda_\ell.
\]
The first step counts the vertices before the last child, in that child, and in $W'\subseteq W$. The second step follows from $\Lambda_{\ell-1}=2^{-t}\Lambda_\ell$. The last step follows from $t\ge512$.

At level zero, the return size is at most $t^3+1\le3\Lambda_0$. The input source bound gives $\Lambda_\ell\ge t|S|$ at every level.

Each full child consumes a nonempty set, and each partial child consumes more than its workload threshold. All loops therefore terminate. At the top level $L:=\lceil\frac{\beta(Q)}{t}\rceil$, the workload exceeds $Q$, so a partial return is impossible and the call finishes fully.
\end{proof}

\subsection{Queue operations and provenance}
\label{sec:queue_provenance}

The same vertex can appear as a source at several levels of the recursion. To count these appearances, we record how each source entry reached the frontier.

\begin{definition}[Provenance tokens]
\label{def:provenance_tokens}
Tokens are namespaced by the current phase. A restart creates fresh phase scratch and tokens while preserving shared labels and cumulative discoveries. No token injection across restarts is asserted. The namespace can be reused after all owned records are cleared.

The following implementation choices remove ambiguity about duplicates and old source records. Attach a token to every source occurrence and queue record. The top source receives a seed token. Explicit insertion from a complete tail $u$ over $(u,v)$ receives token $(u,v)$. Group expansion, representative rescan, and reinsertion of an original source preserve that source's input token. A duplicate insertion keeps the record with smaller stored key and its token. Equality may consistently keep the existing record. These are proof-accounting tokens and require only constant storage per record. Exploratory relaxations do not create tokens in an enclosing engine.
\end{definition}

A source can carry its token through several queues before it is discarded. Following the token through a full call tells us which edge, or which initial seed, pays for the discarded root.

\begin{lemma}[Edge and discarded-root charges]
\label{lem:provenance_injection}
Use the tokens of Definition~\ref{def:provenance_tokens}.
\begin{itemize}
    \item[(a)] Within one engine, each directed edge supplies at most one explicit complete-tail frontier insertion.
    \item[(b)] The discarded-root occurrences at full calls inject into the seed token and these edge tokens. For an engine with $N$ vertices, $M_G$ edges, and levels $0,\ldots,L$, the discarded-root sets $Q_X$ satisfy $\sum_X |Q_X|=O(M_G+1+N(L+1)/t)$, where $X$ ranges over its calls.
\end{itemize}
\end{lemma}
\begin{proof}

{\bf Proof of Part (a).}
We first fix a directed edge and trace the calls that return its tail. Incomparable calls return disjoint sets by Lemma~\ref{thm:capped_return_contract}-(a), so all calls returning its tail form a chain. Completeness fixes the tail label along this chain.

Along this chain, the child handles keys below its input cutoff, and the parent handles keys between that cutoff and its own. The terminal scan is either a base scan below its cutoff or a final $W'$ scan with residual insertion in $[b,B)$. These intervals are disjoint. A level-zero insertion counts as the insertion of its edge. A vertex of $W'$ lies in no child's returned set of the same call, so its final scan is the first scan of that tail in the chain. Hence only one explicit frontier insertion uses the fixed edge.

{\bf Proof of Part (b).}
We next locate the origin tail of a token at a recursive child's input. A parent boundary insertion has tail in $U_i$, with true key below $b_i$, and candidate key at least $B_i\ge b_i$. If the head has true key below $b_i$, exact-prefix or exterior completeness already makes it complete, so this candidate fails the relaxation test.

We obtain the same ordering for a surviving final-$W'$ record and a residual base record: the tail lies below the progress bound and the head lies at or above it. Base records consumed before return never pass to a child.

Since progress cutoffs do not decrease, merges, expansion, and rescanning preserve the fact that the origin tail lies strictly before and outside the current child's target. The seed token has no origin tail and forms a separate charge.

To rule out repeated discarded-root charges, we follow a token when a full call $X$ discards its input source $v$. Every input source lies below the full cutoff, so $v\in U_X$. Its old input token enters no retained group and hence no initial child source set.

To obtain a later occurrence of $v$ below $X$, we need a new explicit boundary insertion. Expansion and source reinsertion only propagate a token already present at their input. The old origin tail lies below the progress of $X$ and belongs to no completion set below $X$, so no descendant recreates the old token. A full return also leaves no original-source residual copy.

We thus rule out a second charge to the same token in both possible call relationships:
\begin{itemize}
\item Incomparable full calls have disjoint returned sets and cannot both contain the token's head.
\item Comparable full calls cannot pass a discarded token from one to the other. Any unpulled copy in a suspended ancestor disappears when the ancestor deletes the returned vertex.
\end{itemize}
Each token is therefore charged by at most one full call. By Part (a) the engine creates at most $M_G$ edge tokens, and it has one seed token. Thus full discarded-root occurrences inject into the seed token and at most $M_G$ edge tokens.

For partial calls, we instead use $|Q_X|\le |S_X|<\frac{|U_X|}{t}$. The returned sets at each level are disjoint, so the partial calls contribute at most $N(L+1)/t$ occurrences in total. Adding the seed and edge charges proves the bound.
\end{proof}

Each recursive call needs a batch of low-key vertices and a cutoff separating them from the rest of the queue. Grouping records into ordered blocks lets us leave the records within each block unsorted.

\begin{definition}[Blocked queue]
\label{def:blocked_queue}
A record stores a vertex,
a frozen scheduling key, a token, and a direct list handle. Maintain an ordered list
of blocks and a deterministic balanced search tree indexing disjoint key
intervals. Block contents are unordered linked lists. Intervals are separated by
saved comparison keys, not by minima that must be recomputed after each deletion.
Use intervals that include their lower endpoint and exclude their upper endpoint. The first block has an unbounded lower endpoint. A pointer identifies it.

Let $M$ be the block parameter.
Ordinary blocks have between $\frac{M}{3}$ and $M$ records. A single sole block may be
smaller. On rebuilding, place blocks in the interior range $[\frac{2M}{5},\frac{4M}{5}]$, with
constant-size rounding adjustments.
\end{definition}

When a child returns a residual queue, its records form a prefix to be merged into its parent's queue. We bound the cost of this merge along with the cost of pulling and updating individual blocks.

\begin{lemma}[Queue bounds]
\label{lem:queue_deletion}
For the blocked queue of Definition~\ref{def:blocked_queue}, let $Q_{\max}$ bound its size and suppose
\[
 \log(2+\frac{Q_{\max}}{M})=O(t),\qquad
 M\geq c\log(2+\frac{Q_{\max}}{M}).
\]
\begin{itemize}
    \item[(a)] Individual insertion costs amortized $O(t)$ time, and deletion by a direct handle costs amortized $O(1)$ time.
    \item[(b)] Pulling the first block takes time linear in its number of records.
    \item[(c)] Merging a separated smaller-block prefix queue takes amortized time linear in its number of records.
    \item[(d)] These amortized bounds hold for every interrupted execution prefix. Base queues can instead be balanced ordered trees.
\end{itemize}
\end{lemma}
\begin{proof}

{\bf Proof of Part (a).}
Every block contains at least $\frac{M}{3}$ records unless it is the sole block, so the interval index has $O(1+\frac{Q_{\max}}{M})$ nodes. We locate an insertion through this index in $O(\log(2+\frac{Q_{\max}}{M}))=O(t)$ time, and unlink a deletion through its direct handle in $O(1)$ time. We still need to pay for repairs when these changes move a block outside its size range.

We implement a decrease or duplicate replacement by a deletion followed by an insertion, and assign it the insertion charge. Every movement updates the handle and preserves one record per vertex.

\paragraph{Step 1: Pay for Block Rebuilds.}
\begin{itemize}
\item On overflow, linear-time selection splits the block into two nearly equal parts.
\item When a block in a queue with multiple blocks falls below $\frac{M}{3}$, join it to a neighbor. Keep the union if it has at most $\frac{4M}{5}$ records. Otherwise split it into two nearly equal parts.
\end{itemize}
In either case we restore the interior size range at cost $O(M+\log(2+\frac{Q_{\max}}{M}))=O(M)$, including handle updates and index changes. Let $T(r)$ denote the worst-case cost of deterministic median-of-five selection on $r$ records.
We have
\begin{equation}
T(r)\le T(\lceil r/5\rceil)+T(7r/10+O(1))+O(r)=O(r).
\label{eq:median_selection}
\end{equation}
The first step follows from selecting a pivot among the group medians and partitioning around it, which leaves at most $7r/10+O(1)$ records in the recursive side. The last step follows by induction on $r$, since $1/5+7/10<1$.

After an interior reset, a block needs $\Omega(M)$ membership changes before another violation. We deposit a sufficiently large constant credit on each change. By Eq.~\eqref{eq:median_selection}, selecting a split among $O(M)$ records costs $O(M)$. The accumulated credit pays for the entire rebuild. Credits on absorbed or removed blocks need not survive.

For bulk creation, we supply the initial credit from the linear record budget. A sole small block needs no underflow repair, its overflow follows $\Omega(M)$ insertions, and removing a sole empty block costs $O(1)$. Thus the bank-account potential stays nonnegative on every prefix. No charge depends on a future extraction.

{\bf Proof of Part (b).}
Consider a pull from the front of the queue.

\paragraph{Step 2: Pull the First Block.}
We pull the first block using the next block's saved lower boundary as separator, or the call's cutoff if no block remains.

In a multiple-block queue, the pulled block has $\Omega(M)$ records. The lower bound on $M$ in Lemma~\ref{lem:queue_deletion} therefore pays for index removal from this linear budget. A small sole block requires only $O(1)$ index work. No pull sorts its records.

{\bf Proof of Part (c).}
Consider a merge into the surviving parent queue.

\paragraph{Step 3: Merge a Separated Prefix.}
Because all incoming keys precede the surviving parent stored keys, we merge at the front. First we remove parent duplicates through their direct handles. The deletion bound charges constant amortized cost per incoming record.

We traverse the child's ordered blocks and concatenate whole blocks until a new prefix block reaches $\frac{2M}{5}$ records. Child blocks are smaller than $\frac{M}{3}$, so the new block remains below $\frac{4M}{5}$, up to the fixed rounding allowance. Its records pay for construction and for the index insertion, which costs $O(\log(2+\frac{Q_{\max}}{M}))=O(M)$ by the lower bound on $M$ in Lemma~\ref{lem:queue_deletion}.

We prepend the final leftover, which has fewer than $\frac{2M}{5}$ records, to the first old parent block without searching the index. That block keeps its upper interval boundary. Any overflow is an ordinary membership-change rebuild.

We only need extra normalization at a small boundary block. If a previously sole small parent becomes part of a multiple-block queue, normalize it with its neighbor. Creating the other block already requires $\Omega(M)$ incoming records, which pays for this work. With an empty parent, apply the same rule to a sole small last block.

We therefore spend $O(r)$ for $r$ incoming records. We dismantle child index trees by traversal, with no repeated balanced-tree deletion. For a base residual tree, traverse in sorted order and batch the records directly.

{\bf Proof of Part (d).}
All credits come from actual record changes or bulk creation. Interruption during a higher-level search preserves these prefix bounds. Creation also prepays linear destruction of remaining allocated or touched records.

\paragraph{Step 4: Check the Engine Parameters.}
To apply these bounds at level $\ell$, we count original sources and out-neighbors of sets completed by that call. These are the only possible queue records. This remains true after a child merge: child inputs come from parent records or group sources, and new child residual records come from the child's returned set.

Bounded degree and the last-child overshoot give $Q_{\max}=O(|S|+\Lambda_\ell)=O(\Lambda_\ell)$.
We have
\begin{equation}
\log(2+Q_{\max}/M_\ell)=O(\log(2+t^2 2^t))=O(t+\log t)=O(t).
\label{eq:queue_index_height}
\end{equation}
The first step follows from $Q_{\max}=O(\Lambda_\ell)$ and the ratio $\Lambda_\ell/M_\ell=t^2 2^t/8$ in Definition~\ref{def:capped_engine}. The second step follows from the logarithm of a product. The last step follows from $\log t\le t$.

Since $M_\ell=8t2^{(\ell-1)t}\ge8t$, the index cost in Eq.~\eqref{eq:queue_index_height} is also $O(M_\ell)$, as required by the block-size hypothesis of Lemma~\ref{lem:queue_deletion}.

For a nonbase child, $M_{\ell-1}=2^{-t}M_\ell<\frac{M_\ell}{3}$, as required by the merge. Base queues have $O(t^3)$ records and $O(\log t)$ update cost. These are the queue operations used by Lemma~\ref{thm:capped_return_contract}-(a).
\end{proof}

\subsection{Parameterized accounting}
\label{sec:parameterized_accounting}

We now combine the cost of the pivot attempts with the cost of maintaining the frontier. Keeping the cost of one attempt explicit lets us use the same estimate at both search scales.

\begin{lemma}[Accounting with explicit attempt coefficients]
\label{lem:explicit_attempt_ledger}
Consider a bounded-degree engine on $N\le Q$ vertices and $M_G\le2Q$ edges, with a supplied size upper bound $Q$. Fix a total observer budget $J$. Suppose direct pivot attempts have charged cost at most $Crf(r)$, where $f\ge1$ is nondecreasing, and parameter evaluation takes $O((J+2)\beta(Q))$ word operations. For a complete phase on at most $Q$ vertices and $M_G\le2Q$ edges, its cost, excluding reusable ambient initialization, is at most
\begin{align*}
K\{&C[Q(L+1)f(k)+(M_G+1)kf(k)]\\
&+(J+2)[M_Gt+Q(L+1)(\frac{t}{k}+1)+Q\log t]+(J+2)\beta(Q)\},
\end{align*}
for an absolute implementation constant $K$, where $L:=\lceil\frac{\beta(Q)}{t}\rceil$, $t\ge512$, and $2\le k\le \frac{t}{256}$. This bound includes repeated local initialization and prepaid owned cleanup, and does not assume $Ckf(k)\le t$.
\end{lemma}
\begin{proof}
We count work through the disjoint completed sets supplied by the return contract, using provenance to control repeated root and edge charges. We then combine these counts with the queue bounds, keeping the pivot coefficient $C$ separate from the current engine's observer factor.

We have
\begin{equation}
\sum_X |U_X|\le N(L+1).
\label{eq:levelwise_returns}
\end{equation}
This bound follows because disjoint returned sets contain at most $N$ vertices in total at each of the $L+1$ levels.

For a full call, Lemma~\ref{lem:target_containment} puts its forest inside $U_X$, and the call returns every input source. For a partial call, $|F_X|=O(k|S_X|)$ and $|U_X|>t|S_X|$, which again give $|F_X|=O(|U_X|)$. These same return bounds control the linear source term in the pivot count by Eq.~\eqref{eq:levelwise_returns}.

Substituting these forest bounds into the pivot estimate of Lemma~\ref{lem:cumulative_forest}-(b), while suppressing $C$ temporarily, gives total pivot work
\begin{align}
& O(N(L+1)f(k)+kf(k)\sum_X|Q_X|+\sum_X|S_X|).
\label{eq:all_pivot_work}
\end{align}
Lemma~\ref{lem:provenance_injection}-(b) charges full calls to the seed and $M_G$ edge tokens, and partial calls to the levelwise returned sets. Applying it here gives
\begin{align}
& \sum_X|Q_X|=O(M_G+1+\frac{N(L+1)}{t}).
\label{eq:discarded_root_charge}
\end{align}
This is the required provenance count. Charging each root to an arbitrary latest predecessor label does not give the injection.

We restore the attempt coefficient $C$ in the pivot bound.
We have
\begin{equation}
\begin{aligned}[b]
\text{pivot work}
&=O(CN(L+1)f(k)+Ckf(k)\sum_X |Q_X|+N(L+1))\\
&=O(CN(L+1)f(k)+Ckf(k)(M_G+1+N(L+1)/t)+N(L+1))\\
&=O(CN(L+1)f(k)+C(M_G+1)kf(k)+N(L+1)).
\end{aligned}
\label{eq:pivot_work_with_coefficient}
\end{equation}
The first step follows from Eq.~\eqref{eq:all_pivot_work}, with the linear source term bounded by Eq.~\eqref{eq:levelwise_returns}. The second step follows from Eq.~\eqref{eq:discarded_root_charge}. The last step follows from $k\le t$, which absorbs the partial-discard term into $CN(L+1)f(k)$.

We also use Lemma~\ref{lem:provenance_injection}-(a) to bound explicit complete-edge frontier insertions by $M_G$. Each costs $O(t)$ by Lemma~\ref{lem:queue_deletion}-(a), so their total is $O(M_Gt)$.

\paragraph{Representative Work.}
We charge $O(\frac{t|U_X|}{k})$ per call for initial representatives. Each rescan and reinsertion costs $O(k+t)=O(t)$. For a partial call there are at most $|S_X|$ representative removals, giving $O(|U_X|)$ total work.

For a full call, we instead count the distinct child sets that remove a current representative of each group. Let $g_j$ be this count for group $j$, and define
\begin{align*}
& E_X^{\mathrm{cross}}:=E(U_X)\setminus\bigcup_i E(U_i).
\end{align*}
The supporting tree $F_j$ contains at least $g_j-1$ edges not wholly contained in any one child completion set.
We have
\begin{equation}
\sum_j g_j\le p_X+|E_X^{\mathrm{cross}}|.
\label{eq:representative_rescans_per_call}
\end{equation}
The inequality follows by summing $g_j-1$ over the $p_X$ supporting trees. Their edge sets are disjoint, so each of the counted edges contributes at most once to $|E_X^{\mathrm{cross}}|$.

To see that a crossing edge receives only one charge, note that its endpoints lie in no single child set. Hence no descendant charges it again. At an ancestor, both endpoints lie in the child containing $X$, so no ancestor charges it either. The full-call crossing charges are therefore globally injective.
We have
\begin{equation}
\sum_{X\text{ full}}\sum_j g_j \le \sum_{X\text{ full}}p_X+\sum_{X\text{ full}}|E_X^{\mathrm{cross}}| \le \sum_{X\text{ full}}p_X+M_G =O(N(L+1)/k+M_G).
\label{eq:representative_rescans_full_calls}
\end{equation}
The first step follows by summing Eq.~\eqref{eq:representative_rescans_per_call} over full calls. The second step follows from the injection of crossing-edge charges into the $M_G$ edges. The last step follows from Lemma~\ref{lem:cumulative_forest}-(c), which gives $p_X=O(|U_X|/k)$, and the levelwise bound in Eq.~\eqref{eq:levelwise_returns}.

We multiply the bound in Eq.~\eqref{eq:representative_rescans_full_calls} by the $O(t)$ cost per rescan and add the partial-call and initial-representative charges. Total representative work is
\begin{align}
& O(M_Gt+\frac{N(L+1)t}{k}+N(L+1)).
\label{eq:representative_charge}
\end{align}

\paragraph{Source and Residual Work.}
We pay for source expansion and original-source reinsertion according to the child return type:
\begin{itemize}
\item A partial child satisfies $|U_i|>t|S_i|$, which absorbs its $O(t|S_i|)$ source charge.
\item A full child consumes its expanded representatives. Charge expansion to their rescan or final group disappearance.
\end{itemize}
A residual queue contains only original sources and out-neighbors of the returned set. Bounded degree gives size $O(|S_i|+|U_i|)$. A full child returns every source, and a partial child has $|S_i|<\frac{|U_i|}{t}$. In either case the residual size is $O(|U_i|)$.

We charge pulls, merges, direct deletion of returned records, and scans to this residual material. The final exhausted-region scan inspects all of $W$. For a full call, Lemma~\ref{lem:target_containment} gives $W\subseteq T(B,S)=U_X$, while the partial case uses $|W|\le k|S_X|<|U_X|$. Both cases give linear scan work in $|U_X|$.

At level zero, disjoint returned sets give total cost $O(N\log t)$, since each queue update costs $O(\log t)$.

The handle and deletion bounds in Lemma~\ref{lem:queue_deletion}-(a), the pull bound in Lemma~\ref{lem:queue_deletion}-(b), and the prefix-merge bound in Lemma~\ref{lem:queue_deletion}-(c) give these charges. Lemma~\ref{lem:queue_deletion}-(d) extends them to every interrupted prefix. A hidden label change alone never changes a stored queue key.

\paragraph{Combine the Charges.}
We combine the pivot work in Eq.~\eqref{eq:pivot_work_with_coefficient} and the representative work in Eq.~\eqref{eq:representative_charge} with the edge, residual, and base charges.

By Lemma~\ref{lem:nested_feasibility}-(c), a current-phase successful relaxation notifies at most $J+2$ environments and monitors. Multiplying current-phase work by $J+2$ covers those notifications. Descendant notifications already belong to $C$, so they receive no second multiplier.

We include owned cleanup by prepaying it at every local creation or initialization, including reuse of an address. Residual transfer moves ownership and clears the old per-vertex handle before installing the new one. Cleanup follows ownership lists and performs no separate balanced-tree deletion for each record. No local phase scans the reusable ambient table.

Finally, we add the assumed $O((J+2)\beta(Q))$ parameter-evaluation cost, replace $N$ by $Q$, and absorb the absolute implementation constants into $K$.
\end{proof}

\subsection{Local interruption}
\label{sec:first_bootstrap}

Algorithm~\ref{alg:capped_attempt} gives the pivot attempt $\textsc{CappedAttempt}_{\mathcal E,c}(x,B,k,F;\lambda)$ for a complete engine $\mathcal E$ and a capacity multiplier $c$. Each phase calls $\mathcal E(x,B,Q;\lambda)$ with the same root $x$ and cutoff $B$ and with advertised capacity $Q=cq$. Thus the capacity changes on a restart, while the root and cutoff remain fixed.

\begin{algorithm}
\caption{The local interruption wrapper}
\label{alg:capped_attempt}
\begin{algorithmic}[1]
\Procedure{CappedAttempt$_{\mathcal E,c}$}{$x,B,k,F;\lambda$} \Comment{Lemma~\ref{prop:interruption}}
\State $C\gets\{x\}$; $E_C\gets\varnothing$; $q\gets2$. Keep $F$ fixed throughout the attempt.
\State Install a monitor that adds each newly discovered vertex and its admitting edge to $C$ and $E_C$.
\State After all notifications, request the first applicable stop: $\mathsf{forest}$ on a forest hit, $\mathsf{size}$ at $|C|=k$, or $\mathsf{capacity}$ at $|C|>q$.
\State Use the checkpoint and outermost-stop priority of Definition~\ref{def:execution_environment}.
\While{true}
    \State $Q\gets cq$. Create fresh phase scratch with only $x$ initially admitted.
    \State Run $\mathcal E(x,B,Q;\lambda)$ until completion or a stop requested by this monitor.
    \State $\sigma\gets$ the stop reason, or $\mathsf{exhausted}$ on completion.
    \State Reclaim the phase's owned scratch and frames using their prepaid deposits.
    \If{$\sigma\ne\mathsf{capacity}$}
        \State \Return $(C,E_C,\sigma)$.
    \EndIf
    \State $q\gets2q$. Retain $C,E_C$, the root seed, and all shared label improvements.
\EndWhile
\EndProcedure
\end{algorithmic}
\end{algorithm}

A complete search has a running-time bound in terms of its advertised capacity. For a pivot attempt, we need a bound in terms of how many vertices it actually discovers before stopping.

\begin{lemma}[Interruption wrapper]
\label{prop:interruption}
Suppose an already proved complete engine $\mathcal E$ solves every admissible
instance of size at most $Q$ in $C Q f(Q)$ time after reusable ambient initialization.
A call $\mathcal E(x,B,Q;\lambda)$ uses root $x$, cutoff $B$, and the current root seed and shared caps. Suppose it has the locality of Lemma~\ref{lem:small_completion} with a fixed degree bound $\Delta=O(1)$, and $f$ is
nondecreasing with $f(C' q)=O(f(q))$ for each fixed $C'$. For $k\ge2$ and $x\notin V(F)$, consider the wrapper $\textnormal{\textsc{CappedAttempt}}_{\mathcal E,c}$ of Algorithm~\ref{alg:capped_attempt}, with capacity multiplier $c:=2(\Delta+1)$.
\begin{itemize}
    \item[(a)] It stops at exhaustion, the first forest hit, or $k$ cumulative discoveries in
$O(rf(r))$ time, where $r$ is the total number of distinct cumulative discoveries.
It returns the cumulative vertices, their first-discovery edges, and the stopping reason of Definition~\ref{def:frontier_targets}.
    \item[(b)] An exhaustion result certifies the entire original feasible region.
\end{itemize}
\end{lemma}
\begin{proof}

{\bf Proof of Part (a).}
A complete-instance bound controls an interrupted prefix only if that prefix also occurs in a sufficiently small admissible instance. We obtain such an instance at each geometric capacity guess and then sum its cost.

At guess $q\in\{2,4,8,\ldots\}$, we use capacity $Q:=2(\Delta+1)q$. The cumulative discovery set $C$ and its first-discovery edges survive restarts. Restart at $|C|>q$ unless a forest hit or the $k$-vertex limit already ends the attempt.

We start each phase with fresh scratch state and only the root admitted, while keeping shared improvements as caps. Lemma~\ref{lem:cap_tightening}-(a)--(b) preserves the feasible problem, and a nonempty return to the root does not improve its seed. Lemma~\ref{lem:nested_feasibility}-(b) preserves every enclosing optimum across these restarts.

To bound a phase prefix, we use that the monitor adds every vertex admitted in the phase to $C$ and requests a restart as soon as $|C|>q$. The prefix therefore has at most $q+1$ phase admissions. A partial return at the top level of the phase would need more than $Q>q+1$ returned vertices, so the monitor stops the phase before one can occur, also when the ambient graph has more than $Q$ vertices. Lemma~\ref{lem:small_completion} reproduces this prefix on an instance with at most $Q$ vertices and the same header. This instance has at most $\Delta(q+1)\le Q$ edges, which meets the edge hypothesis of Lemma~\ref{lem:explicit_attempt_ledger}. The assumed complete bound then gives prefix work $O(qf(q))$.

We reclaim only records that the prefix allocates. Their creation pays for destruction, so cleanup stays within this work bound. The wrapper performs no uncharged scan of ambient arrays.

We double only after more than $q$ distinct cumulative discoveries. Thus the largest guess is at most $2\max\{2,r\}$. The guesses form a geometric sequence and $f$ is nondecreasing, so summing $O(qf(q))$ gives $O(rf(r))$.

{\bf Proof of Part (b).}
To obtain the exhaustion certificate, we apply cap tightening once more: if the final phase exhausts, its feasible reachable set is the original $R$. Every earlier discovery remains in $R$ and appears again in this exhausting phase. Conversely every returned vertex is a cumulative admission. Thus exhaustion returns exactly $C=R$, including discoveries from earlier phases, and supplies the required certificate.
\end{proof}

\subsection{Precomputed orders for short walks}
\label{sec:walk_rank_refinement}

The local searches can share preprocessing of the fixed ambient graph.
We use this to obtain linear-cost base searches, so that no third iterated logarithm enters the bound.
The algorithms below give the complete call chain. \textsc{ShortWalkSSSP} builds the tables once and calls \textsc{WalkBMSSP} with tier $\tau=1$. Each outer pivot attempt runs \textsc{IntermediateAttempt}, whose restart phases call \textsc{IntermediateCapped}. The intermediate engine calls \textsc{WalkBMSSP} with tier $\tau=0$, and its pivot attempts call \textsc{MicroCappedSearch} directly. Ordinary level recursion retains the tier and the shared tables. The micro search has no recursive calls.

We first measure the cost per vertex on the degree-two port graph. The same logarithmic factor will appear in the running time for the original graph.

\begin{definition}[Short-walk running-time profile]
\label{def:short_walk_profiles}
Define
\[
 G(x):=\sqrt{\log x}(\log_2 x)^{1/4},
\]
with logarithms in this expression truncated below by $1$.
\end{definition}

For a fixed root, all paths in a local search share the same inherited prefix. We can therefore order their short suffixes once and reuse that order whenever another search starts at the same root.

\begin{definition}[Ambient short-walk tables]
\label{def:short_walk_tables}
Work on the fixed degree-two port graph $H$, and write $N:=|V(H)|$.
Set
\[
 L:=\beta(N),\qquad
 b:=\lfloor\frac{\lg L}{16}\rfloor,\qquad U:=2^{b+1}-1.
\]
For each root $x$, construct $T_x$ from the empty walk by extending
each node of depth less than $b$ along every outgoing edge in $H$.
Extensions may revisit vertices. The trie has at most $U$ nodes. At each
node store its weight, length, endpoint, parent, and the transition
for each original outgoing slot of its endpoint. Sort its nodes by
\[
 (\text{weight},\text{length},\text{reversed vertex sequence}),
\]
and give them distinct ranks $\operatorname{Rank}_x$ starting at zero. Store the inverse
rank-to-node array $\operatorname{Node}_x$ as well. The empty walk has rank zero. 

Precompute $\operatorname{Pow}[j]:=2^j$ for $0\le j<U$ and an array
$\operatorname{First}[z]$ for $1\le z<2^U$, containing the least
set-bit position of $z$. The entries satisfy
\[
 \operatorname{First}[1]:=0,\qquad
 \operatorname{First}[2z+1]:=0,\qquad
 \operatorname{First}[2z]:=\operatorname{First}[z]+1
\]
whenever the indices lie in the stated range. A set of ranks is represented
by its integer mask. Inserting or deleting a known rank adds or subtracts
its precomputed power of two, and a minimum query uses one array lookup.

Collect the tables and ambient metadata in
\[
 \mathcal T:=(N,b,U,\{T_x,\operatorname{Rank}_x,\operatorname{Node}_x\}_{x\in V(H)},\operatorname{Pow},\operatorname{First}).
\]
Procedure \textsc{BuildShortWalkTables} (Algorithm~\ref{alg:build_walk_tables}) constructs $\mathcal T$ once on $H$. These tables remain read-only.
Every local header retains the ambient value $N$, the tables, and
the original outgoing-slot identifiers. A restricted local graph
keeps those identifiers. The tables are not reconstructed for a
different root, cutoff, inherited history, or capacity guess.
\end{definition}

\begin{algorithm}[!ht]
\caption{Ambient short-walk tables}
\label{alg:build_walk_tables}
\begin{algorithmic}[1]
\Procedure{BuildShortWalkTables}{$H,b$} \Comment{Lemma~\ref{lem:walk_rank_micro}}
\State $N\gets|V(H)|$; $U\gets2^{b+1}-1$.
\For{each $x\in V(H)$}
    \State Initialize $T_x$ with its empty-walk node, of weight $0$, length $0$, and endpoint $x$.
    \For{each trie node $\nu$ of length less than $b$, in breadth-first order}
        \For{each original outgoing slot $(u,v)$ at the endpoint $u$ of $\nu$}
            \State Append a child of weight $\operatorname{weight}(\nu)+w(u,v)$, length $\operatorname{length}(\nu)+1$, endpoint $v$, and parent $\nu$. Store its transition under this slot identifier.
        \EndFor
    \EndFor
    \State Sort the nodes by weight, length, and reversed vertex sequence, in that order.
    \State Assign ranks $\operatorname{Rank}_x$ starting at zero and store the inverse array $\operatorname{Node}_x$.
\EndFor
\State $\operatorname{Pow}[0]\gets1$. Fill $\operatorname{Pow}[j]\gets2\operatorname{Pow}[j-1]$ for $1\le j<U$.
\State $\operatorname{First}[1]\gets0$.
\For{$z$ in $2,\ldots,2^U-1$}
    \If{$z$ is odd}
        \State $\operatorname{First}[z]\gets0$.
    \Else
        \State $\operatorname{First}[z]\gets\operatorname{First}[\frac{z}{2}]+1$.
    \EndIf
\EndFor
\State \Return $\mathcal T$ from Definition~\ref{def:short_walk_tables}.
\EndProcedure
\end{algorithmic}
\end{algorithm}

\begin{algorithm}[!ht]
\caption{A short-walk micro search}
\label{alg:micro_capped_search}
\begin{algorithmic}[1]
\Procedure{MicroCappedSearch}{$x,B,k,F,\mathcal T;\lambda$} \Comment{Lemma~\ref{lem:walk_rank_micro}}
\State Start a local environment with root seed $\lambda(x)$, current caps, and cutoff $B$. Initially admit only $x$ and keep $F$ fixed.
\State Initialize scratch lazily with cleared marks. Reclaim it on every return, preserving $C,E_C$ and shared labels.
\State $C\gets\{x\}$; $E_C\gets\varnothing$; $\pi(x)\gets\operatorname{Node}_x[0]$; $M\gets\operatorname{Pow}[0]$. Mark $x$ queued.
\While{$M\ne0$}
    \State $j\gets\operatorname{First}[M]$; $\nu\gets\operatorname{Node}_x[j]$; $u\gets$ the endpoint of $\nu$.
    \State $M\gets M-\operatorname{Pow}[j]$. Mark $u$ settled and no longer queued.
    \For{each outgoing slot $(u,v)$ of the local graph, with its original identifier}
        \State Form the fresh candidate label $a$ from $\lambda(u)$.
        \If{$\rho(a)<B$ and $a\le\lambda(v)$ under Definition~\ref{def:capped_feasibility}}
            \State $\lambda(v)\gets a$. On first admission, add $v$ to $C$ and $(u,v)$ to $E_C$.
            \State Deliver all admission notifications and dispatch any enclosing stop at the checkpoint of Definition~\ref{def:execution_environment}.
            \If{$v\in V(F)$}
                \State \Return $(C,E_C,\mathsf{forest})$.
            \ElsIf{$|C|=k$}
                \State \Return $(C,E_C,\mathsf{size})$.
            \EndIf
            \If{$v$ is not settled}
                \State $\nu'\gets$ the trie transition from $\pi(u)$ for this original slot.
                \If{$v$ is queued}
                    \State $M\gets M-\operatorname{Pow}[\operatorname{Rank}_x(\pi(v))]$.
                \EndIf
                \State $\pi(v)\gets\nu'$; $M\gets M+\operatorname{Pow}[\operatorname{Rank}_x(\nu')]$. Mark $v$ queued.
            \EndIf
        \EndIf
    \EndFor
\EndWhile
\State \Return $(C,E_C,\mathsf{exhausted})$.
\EndProcedure
\end{algorithmic}
\end{algorithm}

A micro search can use the tables as long as every walk it places in the queue has a precomputed rank. Its admission limit bounds the number of edges in each such walk.

\begin{lemma}[Short-walk micro searches]
\label{lem:walk_rank_micro}
Suppose $b\ge2$ and $2^U\le N$. Then the following hold.
\begin{itemize}
    \item[(a)] Procedure \textnormal{\textsc{BuildShortWalkTables}} (Algorithm~\ref{alg:build_walk_tables}) constructs the tables of Definition~\ref{def:short_walk_tables} in $O(NUb\log(2+U)+2^U)$ time and $O(NU+2^U)$ space.
    \item[(b)] After this preprocessing, procedure \textnormal{\textsc{MicroCappedSearch}} (Algorithm~\ref{alg:micro_capped_search}), with $x\notin V(F)$ and threshold
$2\le k\le b$, has the ordinary exhaustion, size-stop, and first-forest-hit
contracts and costs $O(r)$, where $r$ is its cumulative number of
admitted vertices. Its scratch consists of a suffix pointer $\pi(v)$ and queued/settled marks for each touched vertex, together with the queue mask $M$. The constant is absolute for a fixed observer
budget. The bound includes the triggering operation, callbacks, and
owned cleanup.
\end{itemize}
\end{lemma}
\begin{proof}

{\bf Proof of Part (a).}
Following parent pointers compares two reversed sequences in $O(b)$ time. Sorting the at most $U$ walks at each of $N$ roots therefore costs $O(NUb\log(2+U))$ time and $O(NU)$ space. Weight operations consist only of additions and comparisons.

Computing $\operatorname{First}$ takes $O(2^U)$ time. Since $U=O(L^{1/16})$ and $2^U\le N$, a rank mask fits in one $O(\log N)$-bit metadata word, and the table has size $2^{O(L^{1/16})}=N^{o(1)}$. The algorithm never inspects real-weight bits. These ranks order exact short-walk labels, while the cap test still uses the actual shared label.

{\bf Proof of Part (b).}
To use the precomputed ranks as the capped Dijkstra queue, we fix the specified root $x$ and its inherited seed $\sigma$. Ancestors remain suspended, and their callbacks only record admissions. The seed is fixed: a nonempty suffix returning to the root has nonnegative weight and positive hop count, so the test rejects it. Historical changes before this invocation do not alter the seed's fixed mathematical witness.

At each extraction-loop boundary, the pointer $\pi(v)$ of each queued or settled vertex represents the suffix of its current local witness after $\sigma$, and the mask $M$ contains precisely the queued ranks. Initially the root has the empty-walk pointer. Distinct queued vertices have distinct ranks because their suffixes have different endpoints.

Every queue witness has the same inherited prefix $\sigma$. Its weight and hop count add the same quantities to each suffix, and its reversed identifier sequence follows each reversed suffix. These operations preserve the precomputed ordering. Thus the minimum occupied rank is precisely the minimum full local label. This orders accepted rooted witnesses, not the unrelated inherited caps.

When scanning an edge, the algorithm first performs the ordinary fresh cap and cutoff test. On rejection, it only advances the adjacency iterator.

On acceptance, it writes the shared label and delivers every admission notification before dispatching a stop, as in Definition~\ref{def:execution_environment}. If execution continues and the head is not settled, the original outgoing-slot identifier selects the trie transition from the tail's pointer, and the algorithm updates the head's pointer and queue rank. An accepted equality also installs a fresh local rooted witness, so this update infers no suffix from the old cap's historical predecessor.

A settled vertex is never reinserted. As in ordinary capped Dijkstra, its exact label cannot subsequently decrease: extracted scheduling keys are nondecreasing, and an extension strictly increases the key even across a zero-weight edge. An edge scanned from a later settled tail therefore cannot improve an earlier settled head. These rank comparisons and guarded updates give the ordinary capped Dijkstra induction, so every settled label is its canonical local optimum.

The threshold $k\le b$ also ensures that every needed queued transition lies in the trie. A suffix to an unsettled vertex has no repeated vertex: all preceding vertices are already settled, and nonnegative weights with the hop tiebreak make every extension strictly later than its prefix. Before the $k$th admission, such a suffix and its next extension have at most $k-1$ edges. The $k$th admission stops before its ensuing pointer or queue update.

If the queue exhausts, capped Dijkstra covers the entire feasible region. Otherwise the original size or forest monitor stops the attempt at the prescribed checkpoint. Every accepted write is still a feasible rooted witness, and the first-discovery edges still form the required tree.

Each admitted tail is scanned at most once. Constant degree gives $O(r)$ scans and queue updates, each of constant cost. The attempt has no recursive descendants, and its mask initializes in constant time. Vertex records initialize lazily and enter an ownership list. With a fixed observer budget, notifications and owned cleanup also take $O(r)$ work.

The preprocessing also applies to the restricted instances of Lemma~\ref{lem:small_completion}. Their accepted suffixes remain walks in $H$ with the same order. Unused trie nodes never enter a queue, and the same original slot selects the same transition in the actual and restricted executions.

A rejected probe performs no trie lookup before its guard, and an accepted probe reaches the common post-notification checkpoint before any subsequent queue operation. Retaining the read-only tables in the frozen header therefore preserves the identical execution prefix. Their global construction cost is separate from the local charged work.
\end{proof}

\begin{algorithm}
\caption{The intermediate capped engine}
\label{alg:intermediate_capped}
\begin{algorithmic}[1]
\Procedure{IntermediateCapped}{$x,B,q,\mathcal T;\lambda$} \Comment{Lemma~\ref{lem:walk_rank_intermediate}}
\State Start a fresh local environment with root seed $\lambda(x)$, current caps, and cutoff $B$. Initially admit only $x$.
\State Retain $\mathcal T$ and the original slot identifiers. Use Definitions~\ref{def:capped_feasibility} and~\ref{def:execution_environment} for relaxations and notifications.
\State $t_q\gets512\lceil\sqrt{\beta(q)}\rceil$; $k_q\gets\max\{2,\lfloor\sqrt{\beta(q)}\rfloor\}$.
\If{$t_q>\beta(q)$}
    \State \Return the full feasible set from ordinary capped Dijkstra rooted at $x$.
\EndIf
\State $\ell_q\gets\lceil\frac{\beta(q)}{t_q}\rceil$. Use a phase header of capacity $q$, degree bound $2$, parameters $t_q,k_q$, and top level $\ell_q$.
\State $(b_{\mathrm{out}},U_q,\mathcal D)\gets\Call{WalkBMSSP}{B,\{x\},\ell_q,t_q,k_q,0,\mathcal T;\lambda}$. \Comment{Algorithm~\ref{alg:walk_bmssp}}
\State \Return $U_q$.
\EndProcedure
\end{algorithmic}
\end{algorithm}

The intermediate search uses these linear-time pivot attempts at every recursive level. Its batching parameter and pivot threshold balance the cost of maintaining the queues against the cost of the attempts.

\begin{lemma}[An intermediate engine with shared preprocessing]
\label{lem:walk_rank_intermediate}
Fix the tables of Definition~\ref{def:short_walk_tables} and a constant observer budget, and suppose $2^U\le N$.
\begin{itemize}
    \item[(a)] For every advertised capacity $q$ satisfying
\[
 \max\{2,\lfloor\sqrt{\beta(q)}\rfloor\}\le b,
\]
procedure \textnormal{\textsc{IntermediateCapped}} (Algorithm~\ref{alg:intermediate_capped}) has a complete charged bound
$O(q\sqrt{\beta(q)})$.
    \item[(b)] Its geometric-restart wrapper \textnormal{\textsc{IntermediateAttempt}} (Algorithm~\ref{alg:intermediate_attempt}), with the fixed tables $\mathcal T$ and capacity multiplier $6$, has bound
$O(r\sqrt{\beta(r)})$, provided all of its advertised capacities
satisfy the condition in Part (a).
\end{itemize}
The ambient tables are retained throughout these calls.
\end{lemma}
\begin{proof}

{\bf Proof of Part (a).}
Use the parameters
\begin{align*}
& t_q:=512\lceil\sqrt{\beta(q)}\rceil,\qquad
k_q:=\max\{2,\lfloor\sqrt{\beta(q)}\rfloor\}.
\end{align*}
The hypothesis gives $k_q\le b$, so Lemma~\ref{lem:walk_rank_micro}-(b) supplies linear-cost pivot attempts.

If $t_q>\beta(q)$, the algorithm uses ordinary capped Dijkstra at cost $O(q\beta(q))$. In this case $\beta(q)<512\lceil\sqrt{\beta(q)}\rceil\le1024\sqrt{\beta(q)}$, so $\sqrt{\beta(q)}<1024$ and the cost is $O(q\sqrt{\beta(q)})$ with an absolute constant. The recursive branch below applies only when $\beta(q)\ge512^2$.

Otherwise the parameters satisfy $t_q\ge512$ and $k_q\le\frac{t_q}{256}$. The top call has the single source $x$, which is complete in the fresh environment because Lemma~\ref{cor:restart_without_rollback}-(b) fixes its seed. Its target set is the whole feasible region, so the exterior-completeness invariant of Definition~\ref{def:frontier_targets} holds vacuously. Lemma~\ref{thm:capped_return_contract}-(a) therefore gives correctness of the capped engine. The attempt coefficient is absolute.
We have
\begin{equation}
\text{phase cost}=O(q[t_q+\beta(q)/t_q+\beta(q)/k_q+k_q+\log t_q])
=O(q\sqrt{\beta(q)}).
\label{eq:intermediate_phase_cost}
\end{equation}
The first step follows from Lemma~\ref{lem:explicit_attempt_ledger} with $f=1$, whose remaining terms $\frac{t_q}{k_q}$ and $O(1)$ are at most $t_q$. The last step follows from $t_q,k_q=\Theta(\sqrt{\beta(q)})$, which bounds each term in brackets by $O(\sqrt{\beta(q)})$.

Computing these integer parameters takes $O(\beta(q))$ metadata operations, within the same bound. Ambient preprocessing occurs once, independently of the local capacity $q$.

{\bf Proof of Part (b).}
For the interrupted bound, use the complete-phase cost in Eq.~\eqref{eq:intermediate_phase_cost} and apply Lemma~\ref{prop:interruption}-(a) with $f(q):=\sqrt{\beta(q)}$ and the degree-two capacity multiplier $6$. This profile is nondecreasing and changes by at most an absolute factor under any fixed rescaling. The wrapper hypothesis keeps every advertised capacity in the table-valid range, and Lemma~\ref{prop:interruption} needs the complete bound only at these capacities.

The proof of Lemma~\ref{lem:walk_rank_micro}-(b) shows that small completion preserves the tables and original slot identifiers, so the complete bound applies to comparison instances with the same header. The cumulative discovery set survives restarts, cap tightening preserves the feasible region, and the geometric capacity sum gives $O(r\sqrt{\beta(r)})$. Lemma~\ref{prop:interruption}-(b) gives the exhaustion certificate required by the attempt interface of Definition~\ref{def:frontier_targets}. This application uses the directly established micro-search bound and introduces no circular cost dependence.
\end{proof}

\paragraph{Explicit calls for the two search tiers.}
Algorithms~\ref{alg:walk_pivot_forest} and~\ref{alg:walk_bmssp} give the forest construction and capped engine with the short-walk calls written explicitly. Their tier parameter satisfies $\tau\in\{0,1\}$. For $\tau=0$, require $2\le k\le b$ and use the micro search. For $\tau=1$, require $k\ge2$ and $\max\{2,\lfloor\sqrt{\beta(12k)}\rfloor\}\le b$ and use the intermediate attempt. Both tiers retain the same ambient tables $\mathcal T$ and use parameters satisfying Definition~\ref{def:capped_engine}. The parameter choices in Algorithms~\ref{alg:intermediate_capped} and~\ref{alg:short_walk_sssp} ensure these requirements. Each BMSSP level uses its own queue and workload parameters. Its recursive call changes only the cutoff, source set, and level, retaining $t,k,\tau,\mathcal T$.

The intermediate attempt in Algorithm~\ref{alg:intermediate_attempt} preserves its discovery set across restarts. Each phase calls Algorithm~\ref{alg:intermediate_capped} with the same root and cutoff and with capacity $Q=6q$. The bound $Q\le12k$ ensures the displayed micro-search threshold condition at every phase. Tables and ambient workspace are initialized only by Algorithm~\ref{alg:short_walk_sssp}, not by a local phase.

The forest and queue operations are those of Algorithms~\ref{alg:pivot_forest} and~\ref{alg:bmssp}. Only the pivot-search calls differ. Thus Lemmas~\ref{lem:cumulative_forest}-(a)--(c),~\ref{thm:capped_return_contract}-(a)--(b), and~\ref{lem:explicit_attempt_ledger} apply to the two tiers with their respective attempt bounds. These procedures preserve the notification checkpoints and owned cleanup of Definition~\ref{def:execution_environment}.

\begin{algorithm}
\caption{An intermediate pivot attempt}
\label{alg:intermediate_attempt}
\begin{algorithmic}[1]
\Procedure{IntermediateAttempt}{$x,B,k,F,\mathcal T;\lambda$} \Comment{Lemma~\ref{lem:walk_rank_intermediate}}
\State $C\gets\{x\}$; $E_C\gets\varnothing$; $q\gets2$. Keep $F$ fixed throughout the attempt.
\State Install a monitor that adds each newly discovered vertex and its admitting edge to $C$ and $E_C$.
\State After all notifications, request the first applicable stop: $\mathsf{forest}$ on a forest hit, $\mathsf{size}$ at $|C|=k$, or $\mathsf{capacity}$ at $|C|>q$.
\State Use the checkpoint and outermost-stop priority of Definition~\ref{def:execution_environment}.
\While{true}
    \State $Q\gets6q$. Create fresh phase scratch with only $x$ initially admitted.
    \State Run \Call{IntermediateCapped}{$x,B,Q,\mathcal T;\lambda$} until completion or a stop requested by this monitor. \Comment{Algorithm~\ref{alg:intermediate_capped}}
    \State $\sigma\gets$ the stop reason, or $\mathsf{exhausted}$ on completion.
    \State Reclaim the phase's owned scratch and frames using their prepaid deposits.
    \If{$\sigma\ne\mathsf{capacity}$}
        \State \Return $(C,E_C,\sigma)$.
    \EndIf
    \State $q\gets2q$. Retain $C,E_C$, the root seed, and all shared label improvements.
\EndWhile
\EndProcedure
\end{algorithmic}
\end{algorithm}

\begin{algorithm}[!ht]
\caption{Pivot and forest construction for the two short-walk tiers}
\label{alg:walk_pivot_forest}
\begin{algorithmic}[1]
\Procedure{WalkPivotForest}{$B,S,k,\tau,\mathcal T;\lambda$} \Comment{Lemma~\ref{lem:cumulative_forest}}
\State $F \gets \varnothing$; $Q \gets \varnothing$; $W \gets \varnothing$
\For{each $x\in S$ in the supplied order}
    \If{$x\notin V(F)$}
        \If{$\tau=0$}
            \State $(C,E_C,\sigma)\gets\Call{MicroCappedSearch}{x,B,k,F,\mathcal T;\lambda}$. \Comment{Algorithm~\ref{alg:micro_capped_search}}
        \Else
            \State $(C,E_C,\sigma)\gets\Call{IntermediateAttempt}{x,B,k,F,\mathcal T;\lambda}$. \Comment{Algorithm~\ref{alg:intermediate_attempt}}
        \EndIf
        \If{$\sigma=\mathsf{exhausted}$}
            \State $Q \gets Q\cup\{x\}$; $W \gets W\cup C$
        \Else
            \State $V(F) \gets V(F)\cup C$; $E(F) \gets E(F)\cup E_C$
        \EndIf
    \EndIf
\EndFor
\State $P \gets S\setminus Q$
\State Partition $F$ into supporting trees $F_1,\ldots,F_p$ and assign each $x\in P$ a containing-tree index $j(x)$. Use $p \gets 0$ if $F=\varnothing$.
\State Initialize $P_j \gets \varnothing$ for $1\le j\le p$.
\For{each $x\in P$}
    \State $j \gets j(x)$; $P_j \gets P_j\cup\{x\}$
\EndFor
\State \Return $(\{(P_j,F_j)\}_{j=1}^{p},Q,W)$
\EndProcedure
\end{algorithmic}
\end{algorithm}

\begin{algorithm}[!ht]
\caption{The capped engine for the two short-walk tiers}
\label{alg:walk_bmssp}
\begin{algorithmic}[1]
\Procedure{WalkBMSSP}{$B,S,\ell,t,k,\tau,\mathcal T;\lambda$} \Comment{Lemma~\ref{thm:capped_return_contract}}
\If{$\ell=0$}
    \State Run the level-zero search of Definition~\ref{def:capped_engine}. \Return its triple $(b,U,\mathcal D)$.
\EndIf
\State $(\{(P_j,F_j)\}_{j=1}^{p},Q,W) \gets \Call{WalkPivotForest}{B,S,k,\tau,\mathcal T;\lambda}$ \Comment{Algorithm~\ref{alg:walk_pivot_forest}}
\State Initialize a blocked queue $\mathcal D$ with parameter $M_\ell$; $U\gets\varnothing$.
\For{each nonempty group $P_j$}
    \State Choose a minimum-current-key $p_j\in P_j$; $\alpha_j\gets\rho(\lambda(p_j))$. Insert $(p_j,\alpha_j)$ into $\mathcal D$.
\EndFor
\State $b\gets$ the minimum stored key in $\mathcal D$, or $B$ if empty.
\While{$\mathcal D\ne\varnothing$ and $|U|\le\Lambda_\ell$}
    \State Pull the smallest block $K$ from $\mathcal D$ with separating cutoff $B_i$.
    \State $S_i\gets K$, keeping every pulled record and its token. Use endpoint stamps.
    \State For each current representative in $K$, add its group members with current key below $B_i$ to $S_i$, keeping their input tokens and omitting endpoints already present.
    \State $(b_i,U_i,\mathcal D_i)\gets\Call{WalkBMSSP}{B_i,S_i,\ell-1,t,k,\tau,\mathcal T;\lambda}$ \Comment{Algorithm~\ref{alg:walk_bmssp}}
    \State Delete all vertices of $U_i$ from the parent queue $\mathcal D$, if present. Merge $\mathcal D_i$ into $\mathcal D$.
    \State Mark exactly the groups whose representatives lie in $U_i$. Remove $U_i$ from all groups.
    \For{each outgoing edge $(u,v)$ with $u\in U_i$}
        \State Form its candidate label $\eta$ from the current label of $u$.
        \If{$B_i\le\rho(\eta)<B$ and the shared relaxation succeeds}
            \State Insert $(v,\rho(\lambda(v)))$ into $\mathcal D$.
            \State If $v\in P_j$ for an unmarked group and $\rho(\lambda(v))<\alpha_j$, set $p_j\gets v$ and $\alpha_j\gets\rho(\lambda(v))$.
        \EndIf
    \EndFor
    \State Rescan each marked nonempty $P_j$, choose a minimum-current-key $p_j$, set $\alpha_j\gets\rho(\lambda(p_j))$, and insert $(p_j,\alpha_j)$ into $\mathcal D$.
    \State $U\gets U\cup U_i$; $b\gets b_i$.
\EndWhile
\State If $\mathcal D=\varnothing$, set $b\gets B$.
\State Insert every original source whose current key lies in $[b,B)$ into $\mathcal D$, using its current key and input token.
\State $W'\gets\{v\in W\setminus U:\rho(\lambda(v))<b\}$.
\State Relax outgoing edges of $W'$ below $B$. Insert accepted heads with keys at least $b$ into $\mathcal D$.
\State \Return $(b,U\cup W',\mathcal D)$.
\EndProcedure
\end{algorithmic}
\end{algorithm}

\begin{algorithm}[!ht]
\caption{The short-walk single-source solver}
\label{alg:short_walk_sssp}
\begin{algorithmic}[1]
\Procedure{ShortWalkSSSP}{$G,s_0$} \Comment{Theorem~\ref{thm:short_walk_solver}}
\State Initialize the output distances. Coalesce parallel edges, remove nonnegative self-loops, and restrict to vertices reachable from $s_0$.
\If{the reachable graph has no edges}
    \State \Return distance $0$ at $s_0$ and infinity elsewhere.
\EndIf
\State Apply the degree-two port reduction in the proof of Theorem~\ref{thm:short_walk_solver}. Let $H$ have $N$ vertices and source port $x_0$.
\State $L\gets\beta(N)$; $b\gets\lfloor\frac{\lg L}{16}\rfloor$; $U\gets2^{b+1}-1$.
\State $a\gets\lceil\sqrt{\beta(L)}\rceil$; $p\gets\lceil\sqrt{La}\rceil$; $t\gets512p$; $k\gets\max\{2,\lfloor \frac{L}{p}\rfloor\}$.
\If{$b<\lceil\sqrt{\beta(12L)}\rceil$ or $t>L$ or $U>\lfloor\lg N\rfloor$ or $Ub\lceil\lg(2+U)\rceil>p$}
    \State Run Fibonacci-heap Dijkstra on $H$ from $x_0$. Scatter its port distances and \Return the output array.
\EndIf
\State $\mathcal T\gets\Call{BuildShortWalkTables}{H,b}$. \Comment{Algorithm~\ref{alg:build_walk_tables}}
\State Initialize shared labels with the source label at $x_0$ and infinity elsewhere.
\State Allocate and initialize separate reusable handle columns for the outer engine, intermediate engine, and micro search, with the level envelopes in the proof of Theorem~\ref{thm:short_walk_solver}.
\State Start the outer environment with root seed $\lambda(x_0)$, current caps, and cutoff infinity. Initially admit only $x_0$.
\State $\ell\gets\lceil \frac{L}{t}\rceil$. Freeze the outer header with capacity $N$, degree bound $2$, and parameters $t,k,\ell$.
\State $(b_{\mathrm{out}},R,\mathcal D)\gets\Call{WalkBMSSP}{\infty,\{x_0\},\ell,t,k,1,\mathcal T;\lambda}$. \Comment{Algorithm~\ref{alg:walk_bmssp}}
\State Scatter the numerical port labels to the original reachable vertices and \Return the output array.
\EndProcedure
\end{algorithmic}
\end{algorithm}

The outer search uses the intermediate algorithm for each pivot attempt. We balance the cost of these attempts against the outer queue work and include the one-time cost of building the short-walk tables.

\begin{theorem}[Formal version of Theorem~\ref{thm:informal_main}]
\label{thm:short_walk_solver}
Use $G$ from Definition~\ref{def:short_walk_profiles}. Procedure \textnormal{\textsc{ShortWalkSSSP}}
(Algorithm~\ref{alg:short_walk_sssp}) is a deterministic
comparison-addition algorithm for directed single-source shortest paths
with nonnegative real weights that runs in
$
 O(n+mG(n))
$
time and space, with an absolute implied constant.
\end{theorem}

\begin{proof}
We first analyze Algorithm~\ref{alg:short_walk_sssp} on the fixed port graph $H$, using $N,L,b,U$ from Definition~\ref{def:short_walk_tables}. The outer scale must pay for the intermediate calls as well as the one-time table and handle initialization. Define the integer parameters
\[
a:=\lceil\sqrt{\beta(L)}\rceil,\qquad p:=\lceil\sqrt{La}\rceil,\qquad t:=512p,\qquad k:=\max\{2,\lfloor \frac{L}{p}\rfloor\}.
\]
Before constructing the tables, the algorithm checks
\begin{equation}
b\ge\lceil\sqrt{\beta(12L)}\rceil,\qquad t\le L,\qquad
2^U\le N,\qquad Ub\lceil\lg(2+U)\rceil\le p.
\label{eq:walk_fast_tests}
\end{equation}
It tests these conditions using integer metadata. If $U>\lfloor\lg N\rfloor$, it rejects the fast branch before attempting to represent $2^U$.

All four tests in Eq.~\eqref{eq:walk_fast_tests} pass for every sufficiently large $N$, since $b=\Theta(\log L)$, $U=O(L^{1/16})$, and $p=\Theta(\sqrt L(\log L)^{1/4})$. The ordinary-heap fallback on the remaining fixed finite range therefore has the required asymptotic bound.

We first check that the fast branch supports both search tiers. The degree-two restart wrapper uses advertised capacities at most $12k\le12L$, by the largest-guess bound in Lemma~\ref{prop:interruption}-(a). The first test in Eq.~\eqref{eq:walk_fast_tests} therefore puts every intermediate pivot threshold at most $\lceil\sqrt{\beta(12L)}\rceil\le b$. Also $t\le L$ and $p^2\ge La$ give $2\le k\le\frac{t}{256}$, so the outer parameters satisfy the capped-engine conditions.

We bound the cost of constructing the ambient tables once, using the construction bound of Lemma~\ref{lem:walk_rank_micro}-(a).
We have
\begin{equation}
NUb\log(2+U)=O(Np),\qquad 2^U\le N\le Np.
\label{eq:walk_table_cost}
\end{equation}
The first bound follows from the last test in Eq.~\eqref{eq:walk_fast_tests}, since the two logarithm bases differ by a constant factor. In the second chain, the first step follows from the fast-branch test $2^U\le N$, and the last step follows from $p\ge1$.

With these tables, the algorithm runs \textsc{WalkBMSSP} at tier $\tau=1$ using $t,k$.

Only the outer engine, an intermediate engine, and a nonrecursive micro search are simultaneously active, so the observer budget is an absolute constant. By Lemma~\ref{lem:simple_witnesses}, every stored hop count fits in one metadata word.

By Lemma~\ref{lem:walk_rank_intermediate}-(b), the intermediate interrupted profile is $f(r):=\sqrt{\beta(r)}$. Since $k\le L$, $f(k)\le a$, and $p^2\ge La$, both pivot terms satisfy $kf(k)=O(p)$ and $(\frac{L}{t})f(k)=O(p)$. The condition $t\le L$ also gives $\frac{L}{k}=O(p)$.
We have
\begin{equation}
\begin{aligned}[b]
\text{outer phase cost}
&=O(N[t+(L/t)f(k)+L/k+kf(k)+\log t])\\
&=O(Np)=O(N\sqrt{\log N}(\log_2 N)^{1/4}).
\end{aligned}
\label{eq:walk_outer_cost}
\end{equation}
The first step follows from Lemma~\ref{lem:explicit_attempt_ledger} with the intermediate attempt profile $f$. The second step follows from the two pivot bounds and $L/k=O(p)$, since $t=512p$ also bounds $\log t$ by $O(p)$. The last step follows from $p=\lceil\sqrt{La}\rceil$, with $a=\lceil\sqrt{\beta(L)}\rceil$ and $L=\beta(N)$.

The top-level call has cutoff infinity and the single source $x_0$, which is complete. Every reachable port is then a target with complete ancestor $x_0$, and the exterior-completeness invariant of Definition~\ref{def:frontier_targets} holds vacuously. Lemma~\ref{thm:capped_return_contract}-(a)--(b) therefore returns the complete set $T(\infty,\{x_0\})$, which gives the exact distance of every port. The ports of one original vertex lie on a zero-weight cycle and share that distance.

To include workspace initialization in the same bound, we retain the ambient header and allocate separate reusable columns for the outer and intermediate environments and a constant number for a micro search. The two level counts and the micro search determine the total width.
We have
\begin{equation}
O(\lceil L/t\rceil+\lceil\sqrt{\beta(12L)}\rceil+1)=O(p).
\label{eq:walk_column_count}
\end{equation}
This bound follows by estimating the two level counts separately. The outer count is $O(p)$ because $t=512p$ and $p^2\ge L$. The intermediate count is $O(p)$ because $\beta(12L)=O(L)$.

A fast intermediate phase uses $O(1+\frac{\beta(q)}{t_q})$ levels, and a heap phase needs only a constant number of columns. By Eq.~\eqref{eq:walk_column_count}, the level envelopes and vertex handles initialize once in $O(Np)$ time and space.

At a restart, ownership lists initialize and clear only touched records. There is no table initialization at a local restart. Together with the table costs in Eq.~\eqref{eq:walk_table_cost}, this accounts for all ambient preprocessing within $O(Np)$.

For the general input, we coalesce parallel edges, restrict to the reachable graph, and replace each reachable vertex by one port per incidence, joined in a directed zero-weight cycle. Attach each original edge to its two ports. A one-port gadget needs no loop. This degree-two reduction preserves numerical distances and takes $O(n+m)$ time.

If there are no reachable edges, return directly. Otherwise, for the reachable simple edge count $m'$, the port graph has $N=2m'\le2n^2$. Hence $G(N)=O(G(n))$, and Eq.~\eqref{eq:walk_outer_cost}, together with initialization, reduction, and recovery, gives a running time of $O(n+mG(n))$.

This preprocessing belongs to the fixed ambient port graph. A shortcut graph or an auxiliary graph with a new source requires its own tables.
\end{proof}

%% file: _3_app.tex
\input{49_llm_disclaimer}

%% file: 49_llm_disclaimer.tex
\section*{Acknowledgments and AI Disclosure}

The authors began using AI tools extensively to solve the problem at the end of July 2026. The authors obtained this result on September 29, 2026. Part of the work was done while the second author was visiting Microsoft Research, hosted by Victor Reis. The major idea in this draft comes from multi-level data-structure design and reusing things over iterations in the literature on solving linear programs in matrix multiplication time \cite{cls19}. The AI tools used to solve this problem were ChatGPT 5.6 and Claude Fable 5.0. All of the proofs have been carefully rewritten by the human authors. All of the proofs have been verified by the human authors and different AI models. The authors take full responsibility for the correctness of the proof.

%% file: main.bbl
\newcommand{\etalchar}[1]{$^{#1}$}
\begin{thebibliography}{DMM{\etalchar{+}}25}

\bibitem[AMOT90]{amot90}
Ravindra~K. Ahuja, Kurt Mehlhorn, James~B. Orlin, and Robert~E. Tarjan.
\newblock Faster algorithms for the shortest path problem.
\newblock {\em Journal of the ACM}, 1990.

\bibitem[BCF23]{bcf23}
Karl Bringmann, Alejandro Cassis, and Nick Fischer.
\newblock Negative-weight single-source shortest paths in near-linear time: Now faster!
\newblock In {\em FOCS}, 2023.

\bibitem[Bel58]{bellman58}
Richard Bellman.
\newblock On a routing problem.
\newblock {\em Quarterly of Applied Mathematics}, 16(1):87--90, 1958.

\bibitem[BNWN22]{bnw22}
Aaron Bernstein, Danupon Nanongkai, and Christian Wulff-Nilsen.
\newblock Negative-weight single-source shortest paths in near-linear time.
\newblock In {\em FOCS}, 2022.

\bibitem[CLS19]{cls19}
Michael~B. Cohen, Yin~Tat Lee, and Zhao Song.
\newblock Solving linear programs in the current matrix multiplication time.
\newblock In {\em Proceedings of the 51st Annual ACM SIGACT Symposium on Theory of Computing}, STOC 2019, pages 938--942. Association for Computing Machinery, 2019.

\bibitem[Dij59]{dijkstra59}
Edsger~W. Dijkstra.
\newblock A note on two problems in connexion with graphs.
\newblock {\em Numerische Mathematik}, 1:269--271, 1959.

\bibitem[DMM{\etalchar{+}}25]{dmmsy25}
Ran Duan, Jiayi Mao, Xiao Mao, Xinkai Shu, and Longhui Yin.
\newblock Breaking the sorting barrier for directed single-source shortest paths.
\newblock In {\em ACM Symposium on Theory of Computing (STOC)}, pages 36--44, 2025.

\bibitem[DMSY23]{dmsy23}
Ran Duan, Jiayi Mao, Xinkai Shu, and Longhui Yin.
\newblock A randomized algorithm for single-source shortest path on undirected real-weighted graphs.
\newblock In {\em IEEE Symposium on Foundations of Computer Science (FOCS)}, 2023.

\bibitem[DMSY26]{dmsy26}
Ran Duan, Xiao Mao, Xinkai Shu, and Longhui Yin.
\newblock A faster directed single-source shortest path algorithm.
\newblock In {\em International Colloquium on Automata, Languages, and Programming (ICALP)}, volume 374 of {\em LIPIcs}, pages 81:1--81:23, 2026.

\bibitem[Fin24]{fin24}
Jeremy~T. Fineman.
\newblock Single-source shortest paths with negative real weights in {$\widetilde O(mn^{8/9})$} time.
\newblock In {\em STOC}, 2024.

\bibitem[For56]{ford56}
Lester~R. Ford, Jr.
\newblock Network flow theory.
\newblock Technical Report P-923, RAND Corporation, 1956.

\bibitem[FT87]{ft87}
Michael~L. Fredman and Robert~Endre Tarjan.
\newblock Fibonacci heaps and their uses in improved network optimization algorithms.
\newblock {\em Journal of the ACM}, 34(3):596--615, 1987.

\bibitem[FW94]{fw94}
Michael~L. Fredman and Dan~E. Willard.
\newblock Trans-dichotomous algorithms for minimum spanning trees and shortest paths.
\newblock {\em Journal of Computer and System Sciences}, 1994.

\bibitem[HHR{\etalchar{+}}24]{hhrrt24}
Bernhard Haeupler, Richard Hlad\'{i}k, V\'{a}clav Rozho\v{n}, Robert~E. Tarjan, and Jakub T\v{e}tek.
\newblock Universal optimality of {Dijkstra} via beyond-worst-case heaps.
\newblock In {\em FOCS}, 2024.

\bibitem[HJQ26]{hjq26}
Yufan Huang, Peter Jin, and Kent Quanrud.
\newblock Faster negative length shortest paths by bootstrapping hop reducers.
\newblock In {\em SODA}, 2026.

\bibitem[HLLZ26]{hllz26}
Isaac~M. Hair, George~Z. Li, Jason Li, and Junkai Zhang.
\newblock Bellman-ford in almost-linear time.
\newblock arXiv:2607.19346, 2026.

\bibitem[KR26]{kr26}
Avi Kadria and Liam Roditty.
\newblock A faster undirected single-source shortest path algorithm.
\newblock arXiv:2609.15247, 2026.

\bibitem[PR05]{pr05}
Seth Pettie and Vijaya Ramachandran.
\newblock A shortest path algorithm for real-weighted undirected graphs.
\newblock {\em SIAM Journal on Computing}, 34(6):1398--1431, 2005.

\bibitem[QT26]{qt26}
Kent Quanrud and Navid Tajkhorshid.
\newblock From hop reduction to sparsification for negative length shortest paths.
\newblock In {\em STOC}, 2026.

\bibitem[Tho99]{thorup99}
Mikkel Thorup.
\newblock Undirected single-source shortest paths with positive integer weights in linear time.
\newblock {\em Journal of the ACM}, 46(3):362--394, 1999.

\bibitem[Tho04]{thorup04}
Mikkel Thorup.
\newblock Integer priority queues with decrease key in constant time and the single source shortest paths problem.
\newblock {\em Journal of Computer and System Sciences}, 69(3):330--353, 2004.

\bibitem[Yan26]{yan26}
Shuyi Yan.
\newblock Lossless derandomization for undirected single-source shortest paths and approximate distance oracles.
\newblock In {\em Symposium on Simplicity in Algorithms (SOSA)}, pages 404--412, 2026.

\end{thebibliography}
